%% file: paper.tex
\PassOptionsToPackage{table,xcdraw}{xcolor}
\documentclass[10pt,sigconf,letterpaper,nonacm]{acmart}
\usepackage[nolist]{acronym}
\usepackage[most]{tcolorbox}

\usepackage{bbding,enumitem}
\usepackage{booktabs}

\usepackage{color, soul}
\usepackage{enumitem}

\usepackage{hyperref}
\usepackage{graphicx}

\usepackage{multirow}
\usepackage{pifont}

\usepackage{textcomp}
\usepackage{tikz}
\usepackage{todonotes}

\usepackage{subcaption}
\usepackage{url}
\usepackage{xcolor}

\usepackage[T2A,T1]{fontenc}
\usepackage{xspace}

\usepackage{listings}


\iftrue
\newcommand{\gakiwate}[1]{\textcolor{red}{\noindent[Gautam: #1]}}

\else
\newcommand{\gakiwate}[1]{#1]}

\fi

\newcommand{\ie}{\emph{i.e.}}
\newcommand{\eg}{\emph{e.g.}}
\newcommand{\etal}{\emph{et al.} }
\newcommand{\dns}[1]{{\small \texttt{#1}}}

\newcommand{\urlBiBTeX}[1]{\url{#1}}

\renewcommand{\paragraph}[1]{{\noindent\textbf{{#1}.}\quad}}

\newcommand{\CS}{\emph{Centralization Score}~}
\newcommand{\cs}{$\mathscr{S}$}
\newcommand{\cscc}[1]{$\mathscr{S}_{#1}$}
\newcommand{\csregion}[1]{$\bar{\mathscr{S}}_{#1}$}
\newcommand{\csavg}{$\bar{\mathscr{S}}$}

\definecolor{lightgray}{gray}{0.9}

\usepackage{bold-extra}
\usepackage[tableposition=below]{caption}

\renewcommand\footnotetextcopyrightpermission[1]{}

\copyrightyear{2025}
\acmYear{2025}
\acmConference[SIGCOMM '25]{}{September 8--11, 2025}{Coimbra, Portugal}
\acmBooktitle{ACM SIGCOMM '25, September 8--11, 2025, Coimbra, Portugal}

\begin{document}

\title[On the Centralization and Regionalization of the Web]{On the Centralization and Regionalization of the Web}

\author{Gautam Akiwate}
\affiliation{
    \institution{Stanford University}
    \country{}
}
\email{gakiwate@cs.stanford.edu}

\author{Kimberly Ruth}
\affiliation{
    \institution{Stanford University}
    \country{}
}
\email{kcruth@cs.stanford.edu}

\author{Rumaisa Habib}
\affiliation{
    \institution{Stanford University}
    \country{}
}
\email{rumaisa@cs.stanford.edu}

\author{Zakir Durumeric}
\affiliation{
    \institution{Stanford University}
    \country{}
}
\email{zakir@cs.stanford.edu}

\renewcommand{\shortauthors}{Akiwate, Ruth, Habib, Durumeric}

\begin{abstract}
    \begin{tcolorbox}[colback=red!5!white,colframe=red!75!black]
    An updated and final version of this paper will be published in ACM SIGCOMM '25.
    Please refer to and cite the SIGCOMM version of the paper. The arxiv version
    is now an archival version.
    \end{tcolorbox}
    \input{abstract}
\end{abstract}

\maketitle

\input{figs}
\input{tables}

\input{intro}

\input{background}

\input{centralization}
\input{country_provider}
\input{country_dns}
\input{country_tld}
\input{country_ca}
\input{discussion}
\bibliographystyle{ACM-Reference-Format}
\bibliography{local,misc}

\appendix
\onecolumn
\input{emd_code}
\input{country_references}

\input{emd_scores}
\newpage
\input{cs_all_countries}
\input{insularity_rest}

\end{document}

%% file: abstract.tex
Over the past decade, Internet centralization and its implications for both people and the resilience of the Internet has become a topic of active debate. While the networking community informally agrees on the definition of centralization, we lack a formal metric for quantifying centralization, which limits research beyond descriptive analysis. In this work, we introduce a statistical measure for Internet centralization, which we use to better understand how the web is centralized across four layers of web infrastructure (hosting providers, DNS infrastructure, TLDs, and certificate authorities) in 150~countries. Our work uncovers significant geographical variation, as well as a complex interplay between centralization and sociopolitically driven regionalization. We hope that our work can serve as the foundation for more nuanced analysis to inform this important debate.

%

%% file: figs.tex
\newcommand{\emdscoreexample}{
\begin{figure*}[t!]
    \centering
    \includegraphics[width=0.9\linewidth]{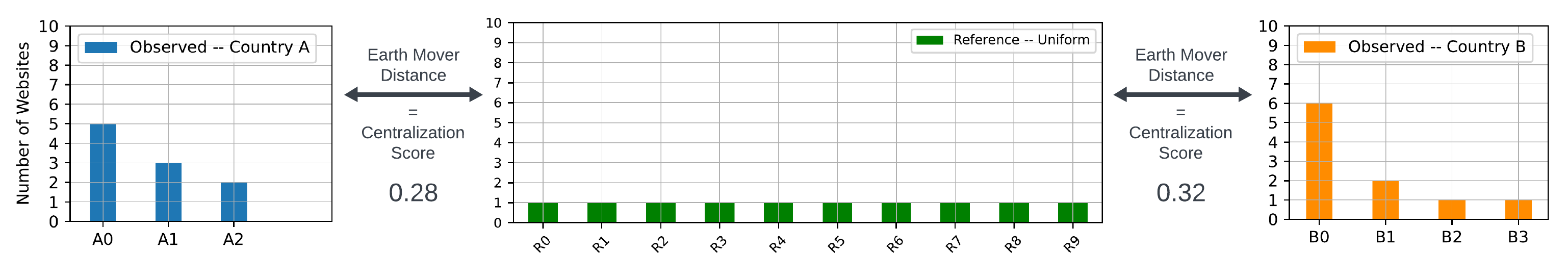}
    \vspace{-8pt}
    \caption{Centralization Comparison Example---\textnormal{To calculate the
    centralization score for the top websites in Countries A and B,
    we calculate the EMD between the observed
    distribution in each to a reference uniform distribution. 
    In the example
    above, the EMD for Countries A and B are 0.28 and 0.32, respectively, indicating that Country A is \emph{less centralized} than B.
    }}
    \label{fig:centralization_example}
\end{figure*}
}

\newcommand{\cdfcountryexample}{
\begin{figure}[t]
    \centering
    \includegraphics[width=\columnwidth]{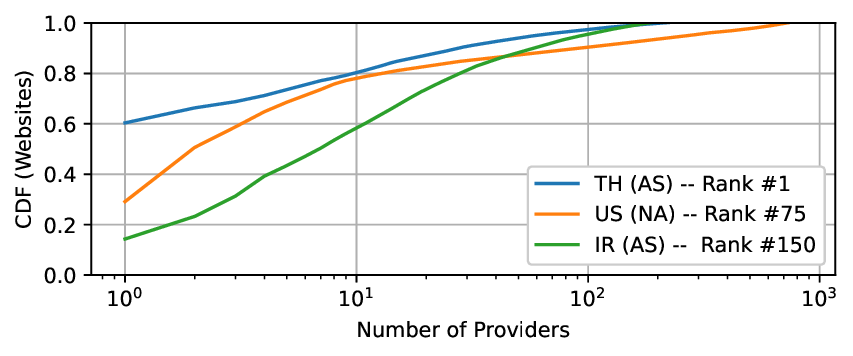}
    \vspace{-15pt}
    \caption{CDF of Hosting Providers---\textnormal{Differences in centralization are largely driven by the distribution of sites amongst the ten largest providers in each country. For both the most centralized (Thailand) and least centralized (Iran) countries, 100~providers account for most sites.}}
    \label{fig:centralization_country_cdf_examples}
\end{figure}
}

\newcommand{\cshistogramall}{
\begin{figure*}[tp]
    \centering
    \begin{subfigure}[t]{0.24\linewidth}
        \centering
        \includegraphics[width=\linewidth]{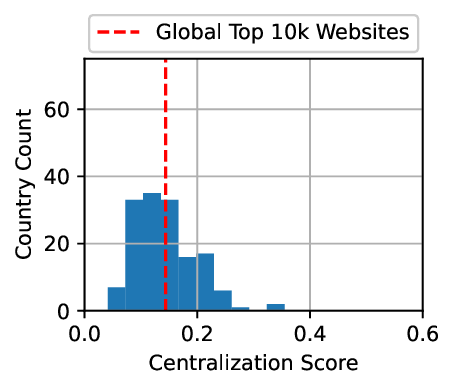}
        \caption{Hosting Providers}
        \label{fig:centralization_score_histogram_hosting}
    \end{subfigure}%
    \begin{subfigure}[t]{0.24\linewidth}
        \centering
        \includegraphics[width=\linewidth]{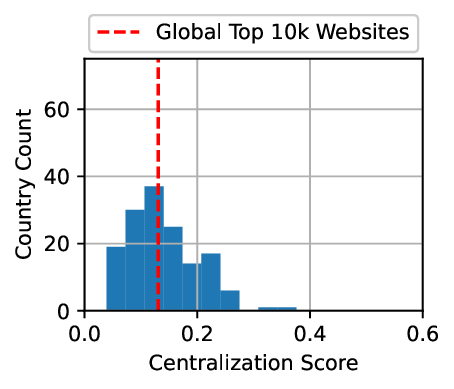}
        \caption{DNS Providers}
        \label{fig:centralization_score_histogram_dns}
    \end{subfigure}%
   \begin{subfigure}[t]{0.24\linewidth}
        \centering
        \includegraphics[width=\linewidth]{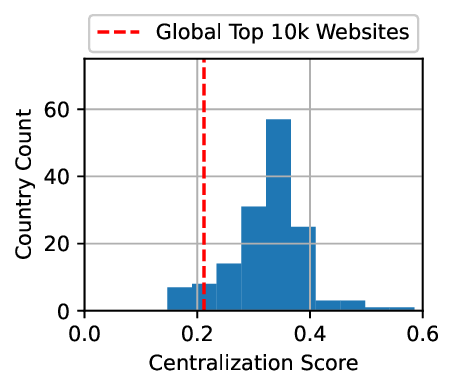}
        \caption{TLDs}
        \label{fig:centralization_score_histogram_tld}
    \end{subfigure}
    \begin{subfigure}[t]{0.24\linewidth}
        \centering
        \includegraphics[width=\linewidth]{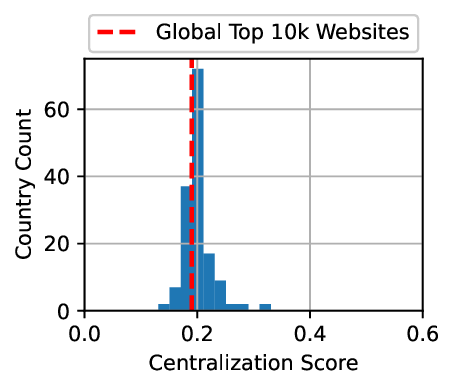}
        \caption{Certificate Authorities}
        \label{fig:centralization_score_histogram_cas}
    \end{subfigure}
    \caption{Centralization Distributions By Country---\textnormal{The distributions for hosting and
    DNS providers are roughly similar. In contrast, the distribution for CAs has
    very little variance given the smaller number of providers and dominant use of large
    global CAs. The histogram for TLDs shows higher centralization scores all across,
    indicating that countries depend on few TLDs. Also marked is \cs~ for
    the Global Top 10k websites and shows that while it is representative of the average centralization
    in hosting, DNS, and CA layers, it is not representative for TLDs.}}
    \label{fig:centralization_score_histograms}
\end{figure*}
}

\newcommand{\cscontinentboxplot}{
\begin{figure*}[tp]
    \centering
    \begin{subfigure}[t]{0.24\linewidth}
        \centering
        \includegraphics[width=\linewidth]{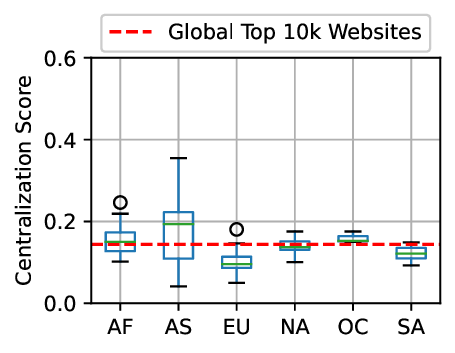}
        \caption{Hosting Providers}
        \label{fig:centralization_score_boxplot_hosting}
    \end{subfigure}%
    \begin{subfigure}[t]{0.24\linewidth}
        \centering
        \includegraphics[width=\linewidth]{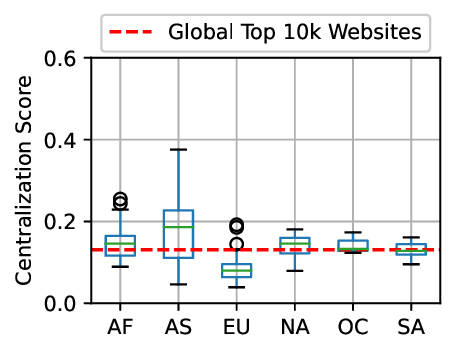}
        \caption{DNS Providers}
        \label{fig:centralization_score_boxplot_dns}
    \end{subfigure}%
    \begin{subfigure}[t]{0.24\linewidth}
        \centering
        \includegraphics[width=\linewidth]{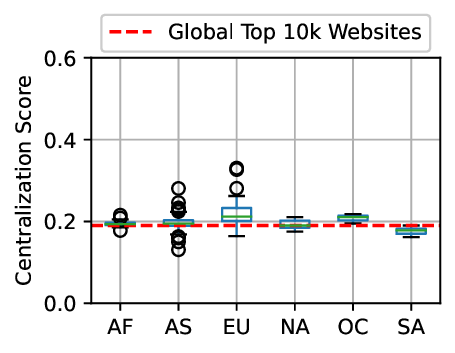}
        \caption{Certificate Authorities}
        \label{fig:centralization_score_boxplot_ca}
    \end{subfigure}
    \begin{subfigure}[t]{0.24\linewidth}
        \centering
        \includegraphics[width=\linewidth]{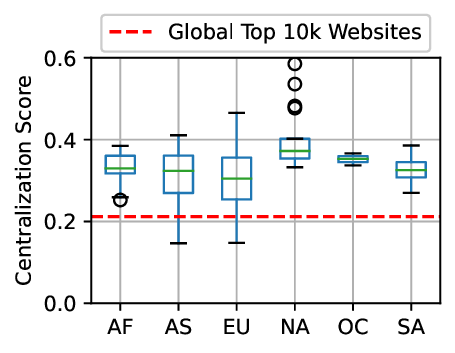}
        \caption{TLDs}
        \label{fig:centralization_score_boxplot_tld}
    \end{subfigure}
    \caption{Boxplot illustrating the distribution of Centralization Scores (CS)
    for countries in different continents: AF (Africe, N=34), AS (Asia, N=46), EU
    (Europe, N=40), NA (North America, N=17), OC (Oceania, N=3), and SA (South
    America, N=10). The dashed red line shows the CS for the Global Top10K websites. We
    see Europe is consistently least centralized for hosting and DNS while skewing towards
    being the most centralized for CAs.} 
    \label{fig:centralization_score_boxplots}
    \end{figure*}
}

\newcommand{\interpguide}{
\begin{figure}[t]
    \centering
    \includegraphics[width=0.95\columnwidth]{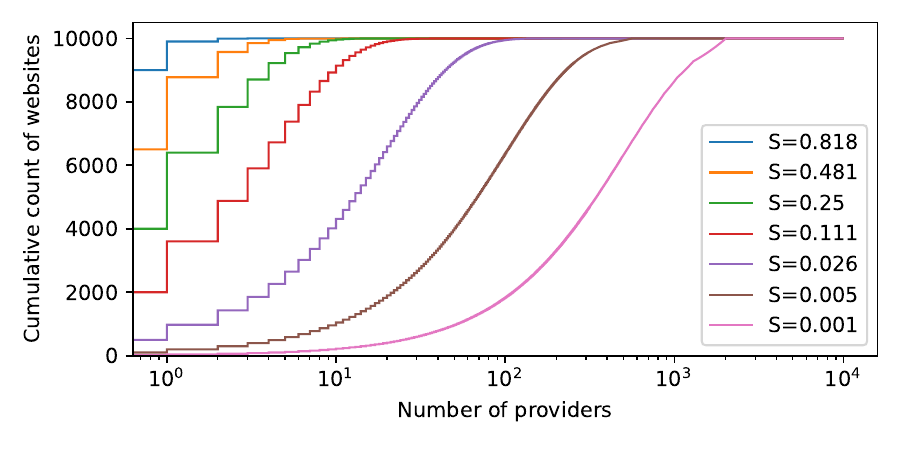}
    \vspace{-10pt}
    \caption{Example \cs\xspace Values---\textnormal{Centralization Score (\cs) for multiple synthetic distributions. \cs\xspace values are most sensitive to differences between the highly centralized cases.
    }}
    \label{fig:interpretation_guide}
    \vspace{-5pt}
\end{figure}
}

\newcommand{\usagecurve}{
\begin{figure}[t]
    \centering
    \includegraphics[width=0.8\columnwidth]{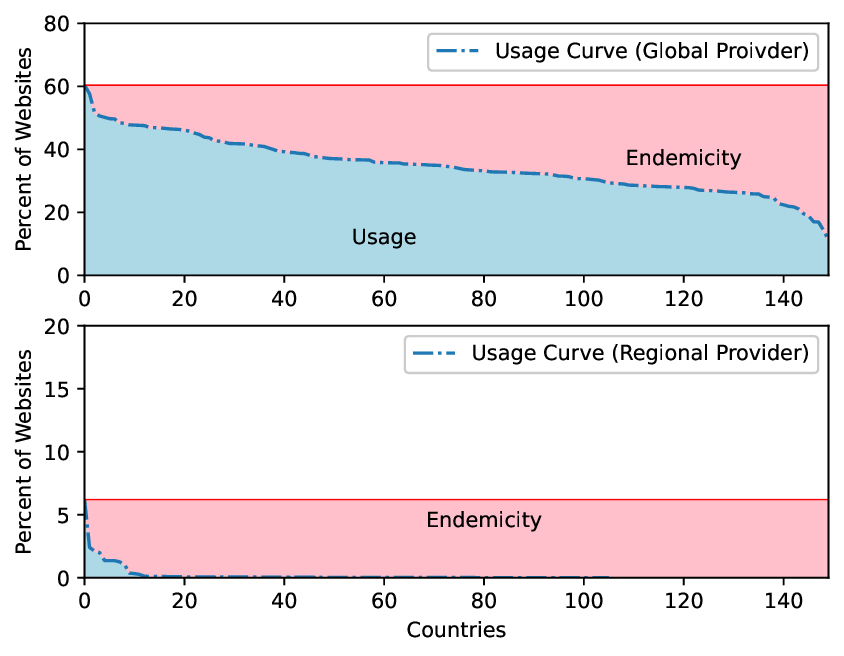}
    \vspace{-7pt}
    \caption{Usage and Endemicity---\textnormal{Usage ($U$) is the area under the usage curve and endemicity ($\mathcal{E}$) is the area between the usage curve and
    the horizontal line starting at the usage curve's maximum value. Usage captures popularity, while endemicity captures global consistency in usage. Regional providers have a higher endemicity
    than global providers.}
    } 
    \label{fig:provider_usage_curve}
\end{figure}
}

\newcommand{\hostingclusters}{
\begin{figure}[t]
    \centering
    \includegraphics[width=0.9\columnwidth]{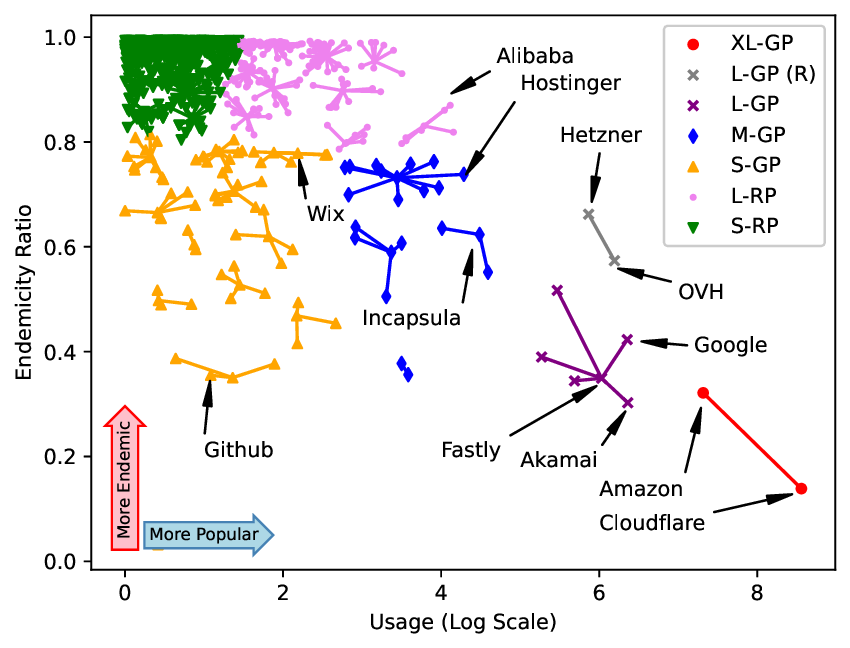}
    \caption{Classification of Providers---\textnormal{We cluster
    and classify providers based on two dimensions: their size and their
    endemicity. We visualize 7 classes, with XS-RP class not visualized for clarity.
    L-GP indicates a large (based on use in the top websites) global provider while
    L-RP indicates a large regional provider (based on endemicity).}}
    \label{fig:categorizing_providers_clusters}
\end{figure}
}

\newcommand{\hostingheatmapcc}{
\begin{figure}[t]
    \centering
    \includegraphics[width=\columnwidth]{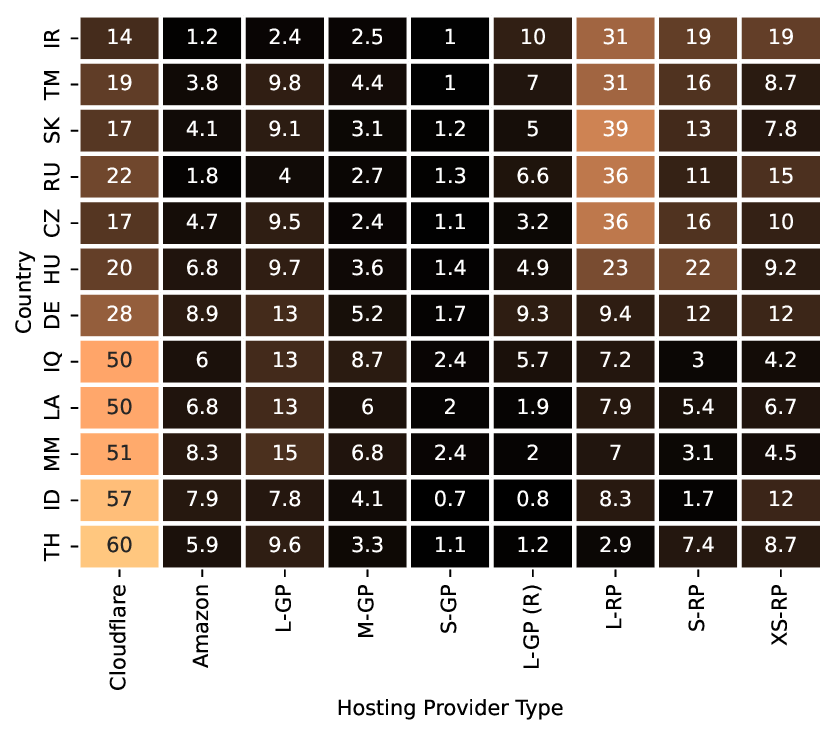}
    \caption{Percentage of websites in the Top 5 most and least centralized
    countries broken down by provider type. The least centralized countries tend
    to rely on regional providers while the most centralized countries. \gakiwate{Remove after writing text.}}
    \label{fig:heatmap_providers_country}
\end{figure}
}

\newcommand{\orggeo}{
\begin{figure}[t]
    \centering
    \includegraphics[width=0.9\columnwidth]{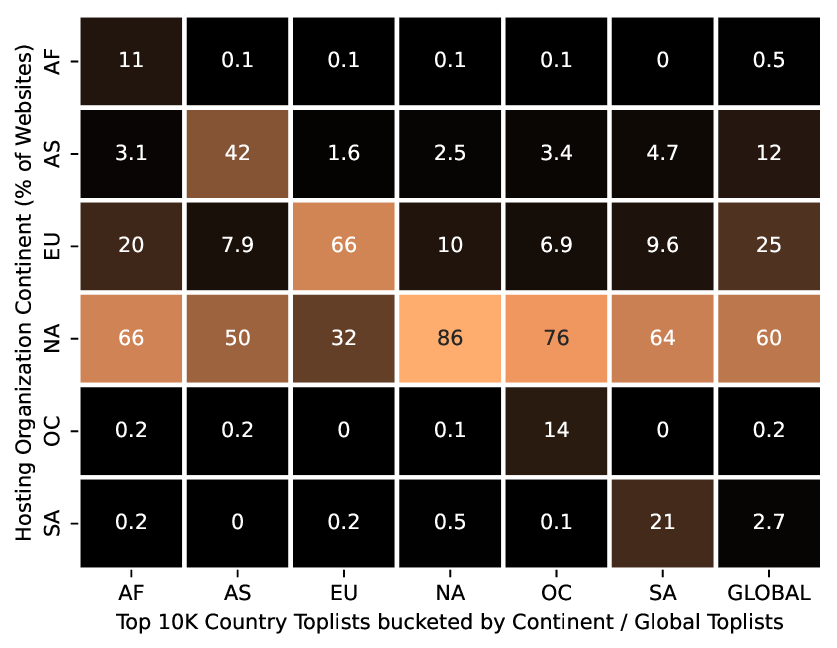}
    \caption{Percentage of websites in the top websites grouped by continent
    broken down by the continent in which the hosting provider is located.}
    \label{fig:hosting_org_continent}
\end{figure}
}

\newcommand{\orggeoregion}{
\begin{figure*}[t]
    \centering
    \centering
    \begin{subfigure}[t]{0.32\linewidth}
    \includegraphics[width=\columnwidth]{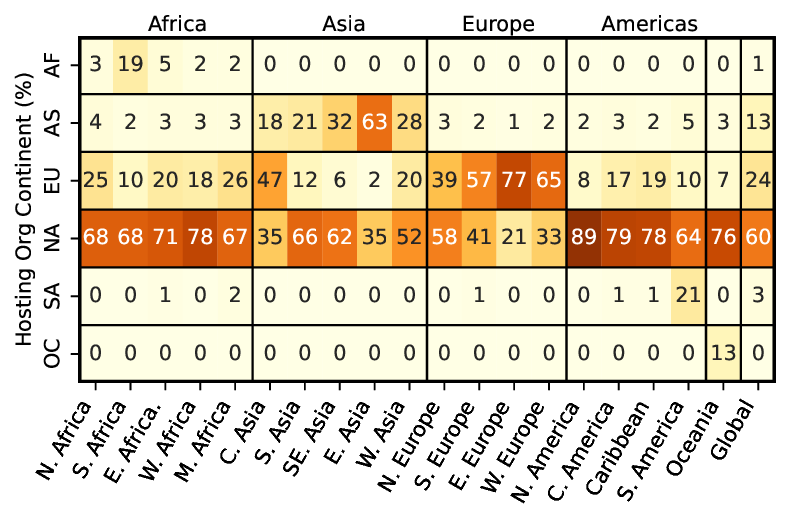}
    \caption{Hosting Provider H.Q.}
    \label{fig:hosting_region_org_continent}
    \end{subfigure}
    \begin{subfigure}[t]{0.32\linewidth}
    \centering
    \includegraphics[width=\columnwidth]{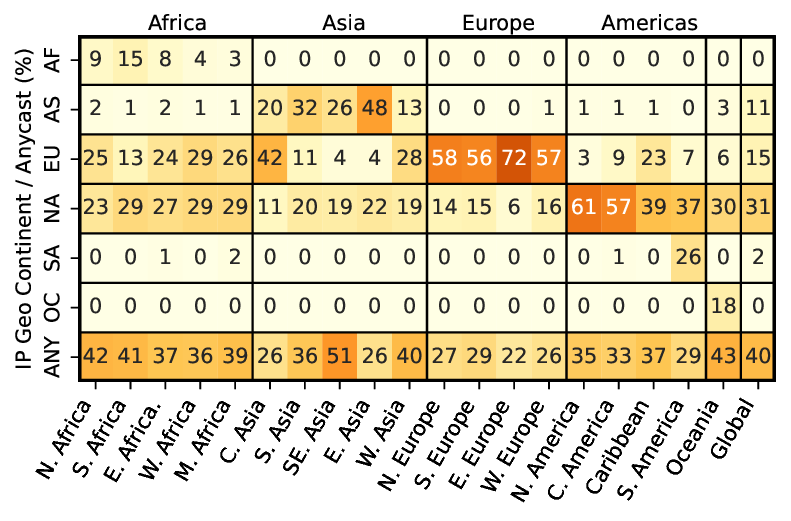}
    \caption{IP Geolocation}
    \label{fig:ip_geo_region_continent}
    \end{subfigure}
    \begin{subfigure}[t]{0.32\linewidth}
        \centering
    \includegraphics[width=\columnwidth]{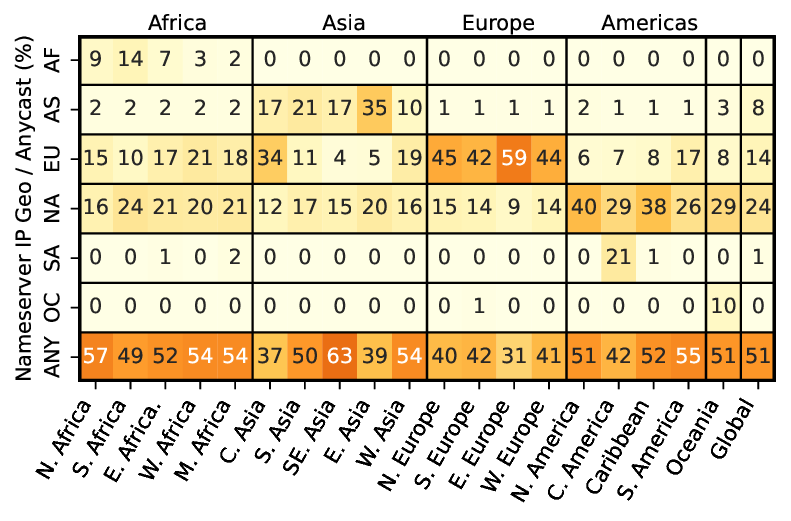}
    \caption{DNS Nameserver Geolocation}
    \label{fig:dns_geo_region_continent}
    \end{subfigure}
    \caption{Regional Dependencies on Other Continents---\textnormal{Since most global providers are headquartered
    in the US, we see a strong reliance on N. America across the board. This mostly shifts into anycast 
    looking at IP geolocation since most global providers (\eg, Cloudflare) operate anycast prefixes.
    Note, we see an increased shift to anycast prefixes when looking at DNS. While Europe and Eastern Asia (Japan, Korea)
    are mostly self-reliant, Central Asia shows use of providers and services in Europe (Russia). Africa primarily
    relies on providers and services in North America and Europe.}}
    \label{fig:hosting_region_geo_continent}
\end{figure*}
}

\newcommand{\ipgeo}{
\begin{figure}[t]
    \centering
    \includegraphics[width=0.9\columnwidth]{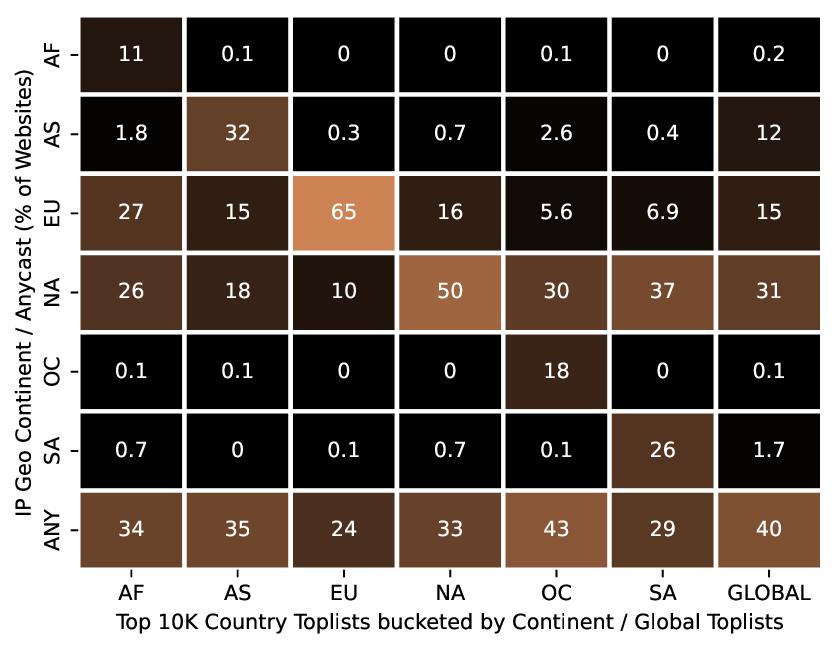}
    \caption{Percentage of websites in the top websites grouped by continent
    broken down by the continent in which the IP is geolocated. Anycast IPs are
    considered as a separate category.}
    \label{fig:hosting_geo_continent}
\end{figure}
}

\newcommand{\csproviderbarplot}{
 \begin{figure*}[t]
    \centering
    \includegraphics[width=\linewidth]{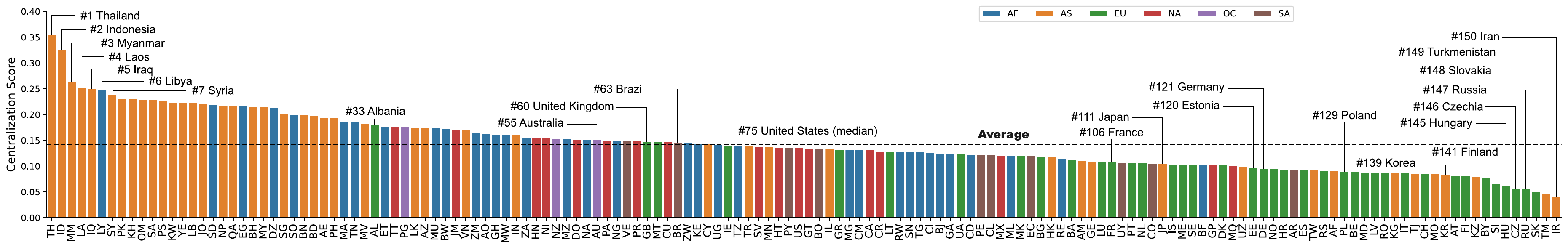}
    \caption{Hosting Provider Centralization by Country---\textnormal{Europe is consistently the least centralized, 
    while Asia as a whole shows a lot of variance. Other continents do not tend towards any extremes.}}
    \label{fig:centralization_barplot}
\end{figure*} 
}

\newcommand{\csdnsbarplot}{
 \begin{figure*}[t]
    \centering
    \includegraphics[width=.95\linewidth]{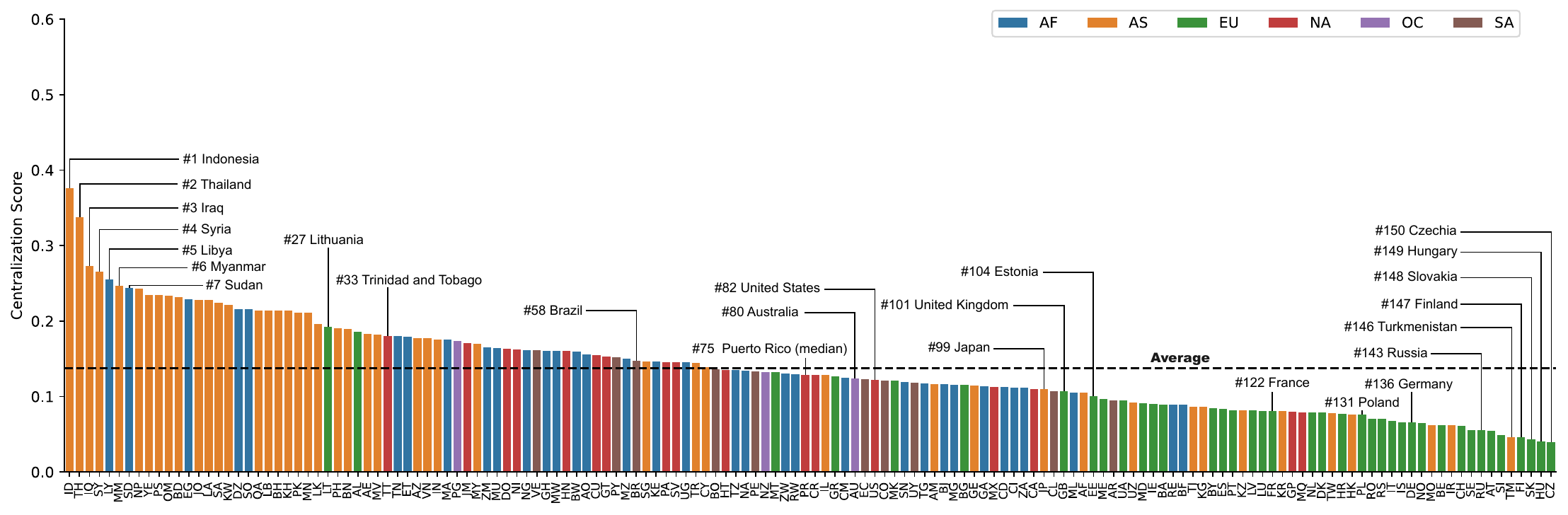}
    \caption{DNS Centralization (sorted by \cs, color coded by continent)---\textnormal {It shows similar trends to that of hosting providers \ie~ European countries tend to be less centralized, while Southeast Asian countries tend to be more centralized.}}
    \label{fig:centralization_dns_barplot}
\end{figure*} 
}

\newcommand{\cscabarplot}{
 \begin{figure*}[t]
    \centering
    \includegraphics[width=.95\linewidth]{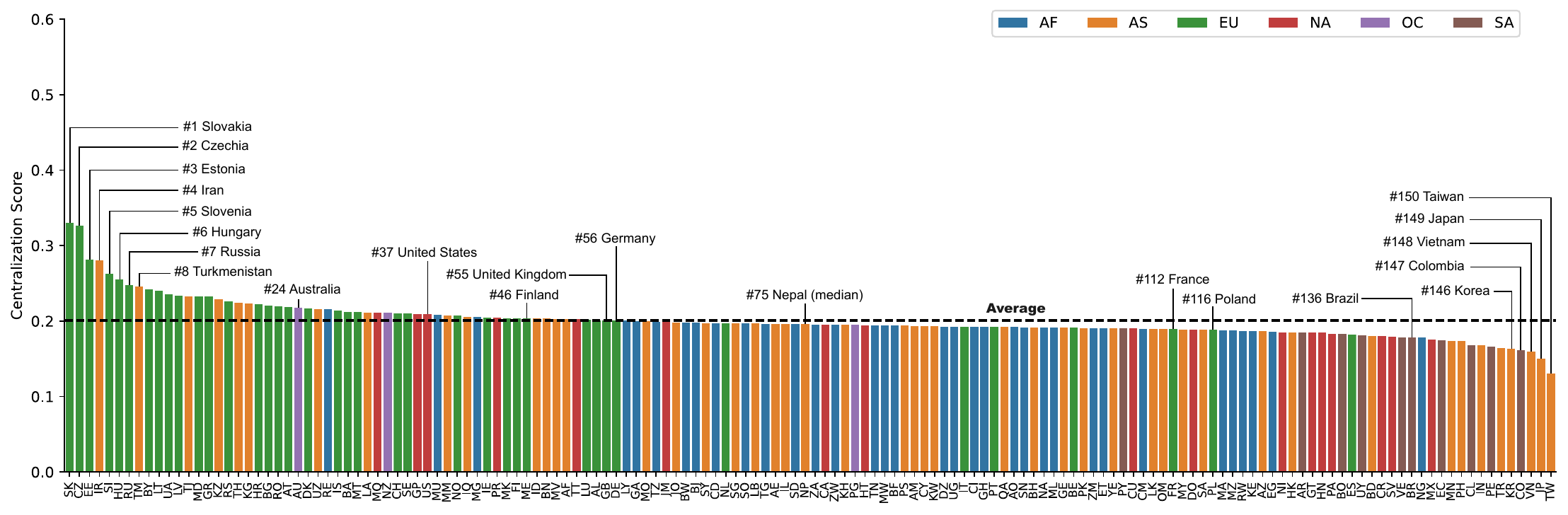}
    \caption{CA Centralization (sorted by \cs, color coded by continent)---\textnormal{We see a shift from the hosting and DNS patterns. European countries tend to be more centralized in this layer.}}
    \label{fig:centralization_ca_barplot}
\end{figure*} 
}

\newcommand{\cstldbarplot}{
 \begin{figure*}[t]
    \centering
    \includegraphics[width=.95\linewidth]{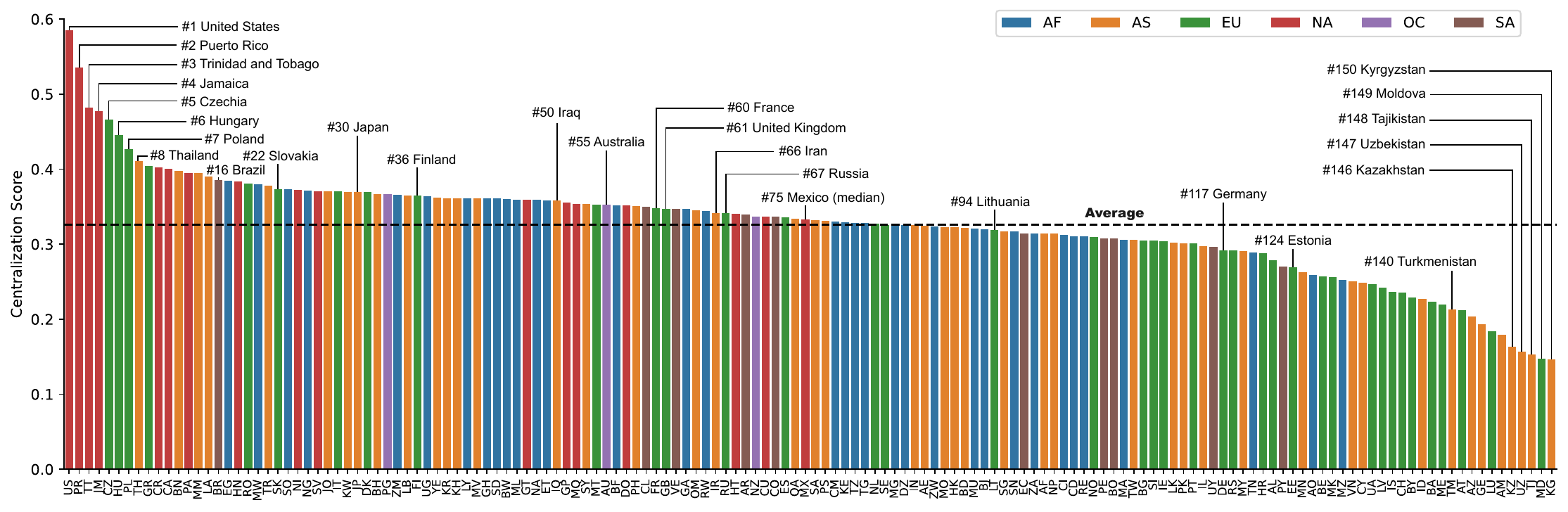}
    \caption{TLD Centralization (sorted by \cs, color coded by continent)---\textnormal{North American countries show a tendency to be centralized. The CIS countries are on the other extreme, due to their dependence on a TLD apart from their own (\dns{.ru}) in addition to the globally popular ones.}}
    \label{fig:centralization_tld_barplot}
\end{figure*} 
}

\newcommand{\providerstackedbarplot}{
 \begin{figure*}[t]
    \centering
    \includegraphics[width=\linewidth]{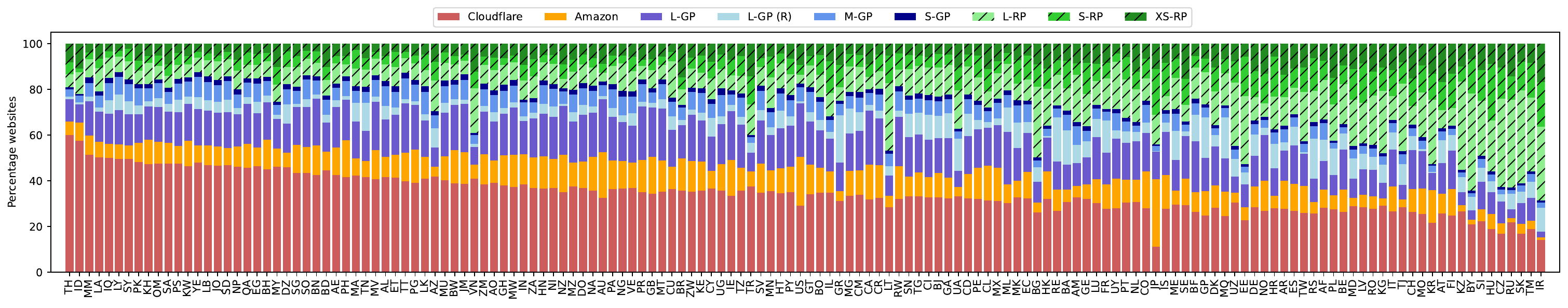}
    \caption{Breakdown of Hosting Provider Types Across Countries---\textnormal{Cloudflare is the most popular provider in every country except Japan; while the most centralized countries
    overtly rely on Cloudflare, the least centralized countries tend to rely on a range of regional providers (hatched bars).}}
    \label{fig:hosting_provider_stacked}
\end{figure*} 
}

\newcommand{\dnsstackedbarplot}{
 \begin{figure*}[t]
    \centering
    \includegraphics[width=\linewidth]{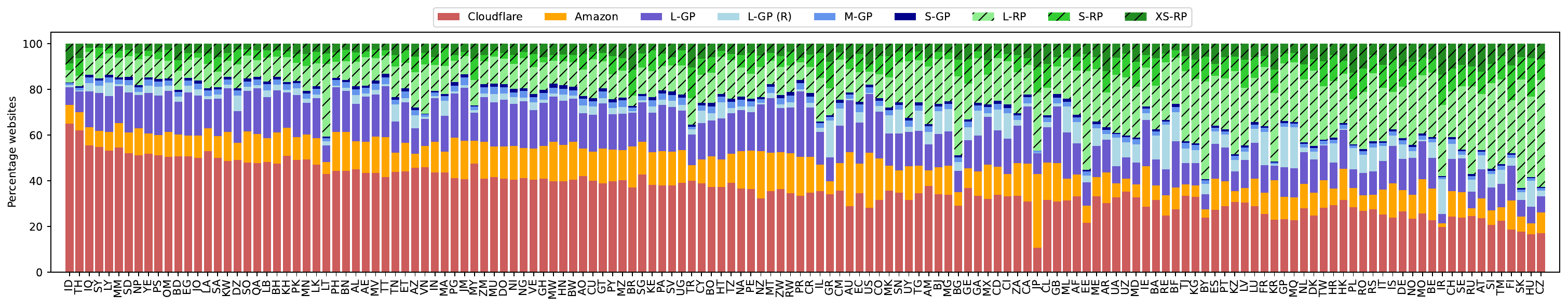}
    \caption{Percentage of websites in all the countries (sorted by \cs) broken down by DNS provider type---\textnormal{Similar to the hosting provider case, Cloudflare dominates in every country save for Japan.}}
    \label{fig:hosting_dns_stacked}
\end{figure*} 
}

\newcommand{\castackedbarplot}{
 \begin{figure*}[t]
    \centering
    \includegraphics[width=\linewidth]{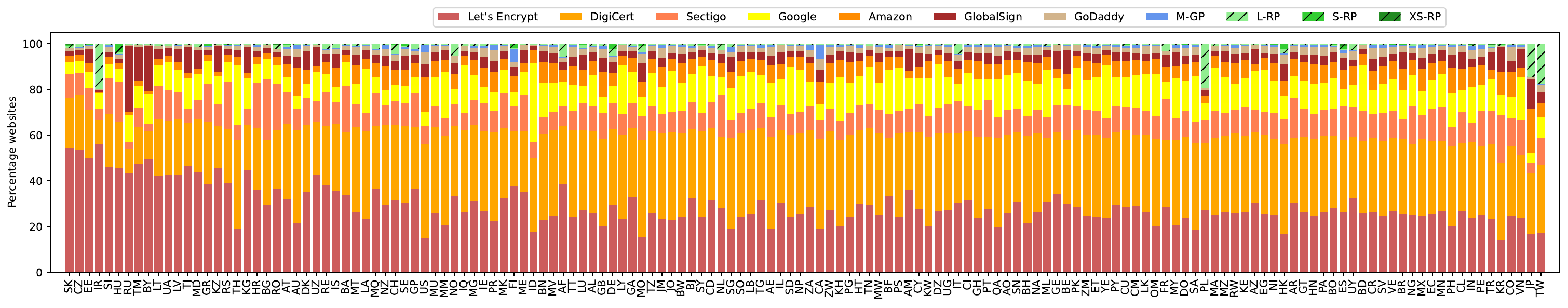}
    \caption{Percentage of websites in all the countries (sorted by \cs) broken down by CA type---\textnormal{Notably, the 7 large CAs account for an average of 98\% of websites across countries.}}
    \label{fig:hosting_ca_stacked}
\end{figure*} 
}

\newcommand{\tldstackedbarplot}{
 \begin{figure*}[t]
    \centering
    \includegraphics[width=\linewidth]{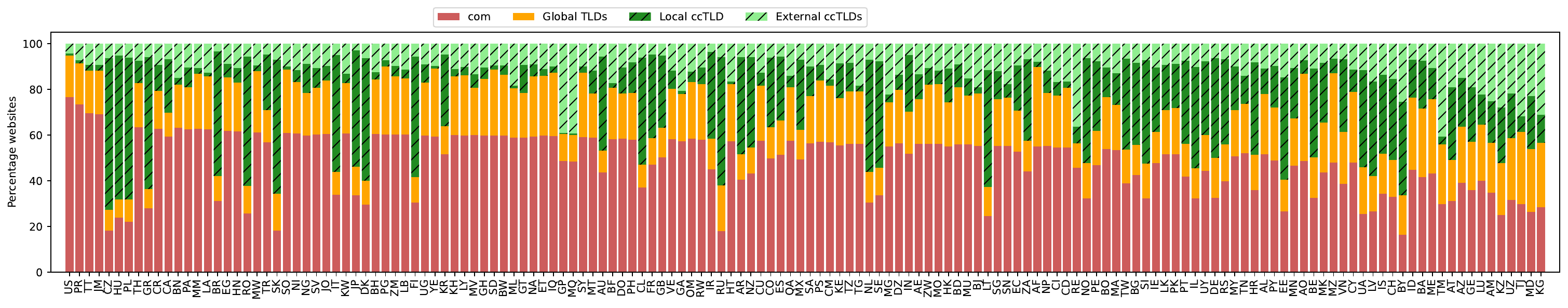}
    \caption{Percentage of websites in all the countries (sorted by \cs) broken down by TLD type---\textnormal{The usage of external ccTLDs is strongly correlated with lower centralization.}}
    \label{fig:hosting_tld_stacked}
\end{figure*} 
}

\newcommand{\csregionboxplot}{
 \begin{figure*}[t]
    \centering
    \includegraphics[width=\linewidth]{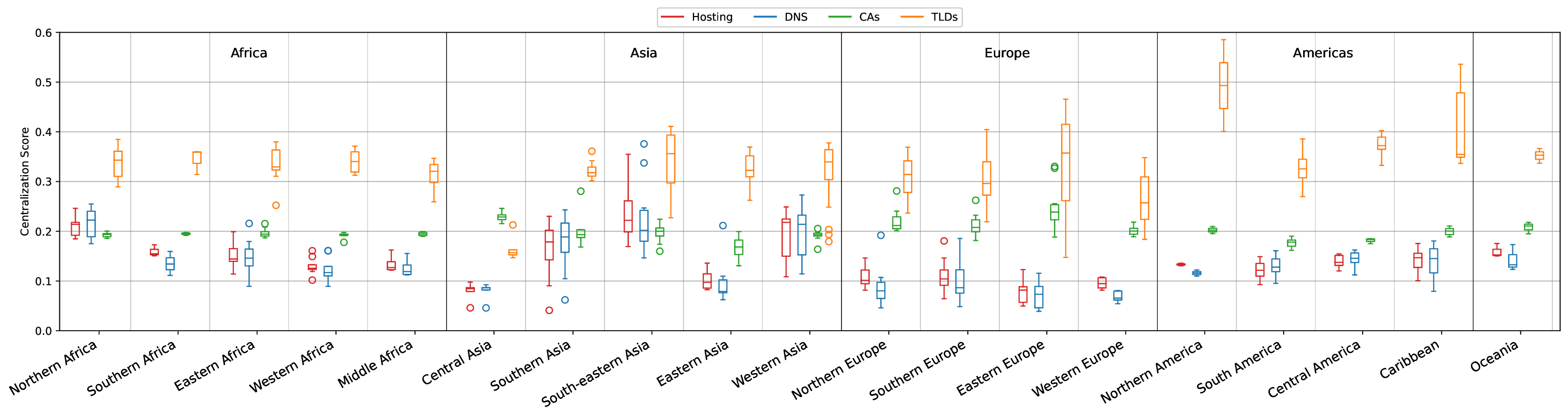}
    \caption{Centralization Across Layers and Subregions---\textnormal{Distribution of \cs~ across the hosting and DNS layers look roughly similar, while CA usage shows minimal variance in centralization given the limited number of CAs, and TLD use shows a higher degree of centralization and variance compared to other layers. 
    }} 
    \label{fig:cs_boxplots_regions}
\end{figure*}
}

\newcommand{\insularityboxplot}{
 \begin{figure*}[t]
    \centering
    \includegraphics[width=\linewidth]{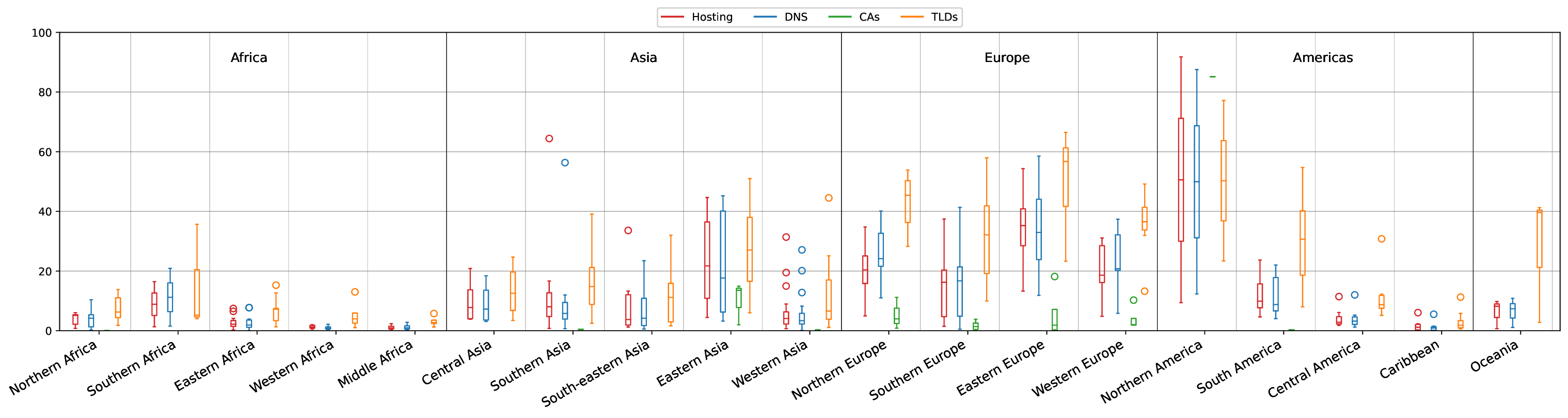}
    \caption{Insularity Across Layers and Subregions---\textnormal{Most global providers are based in the U.S., which drives North America to be the most insular region. Countries in Europe and Eastern Asia are consistently the most insular countries across all layers. Countries in the Global South show insularity at the TLD layer, but have low insularity in other layers since equivalent providers do not exist locally.}}
    \label{fig:insularity_boxplots_regions}
\end{figure*}
}

\newcommand{\cdfinsularity}{
\begin{figure}[t]
    \centering
    \includegraphics[width=\columnwidth]{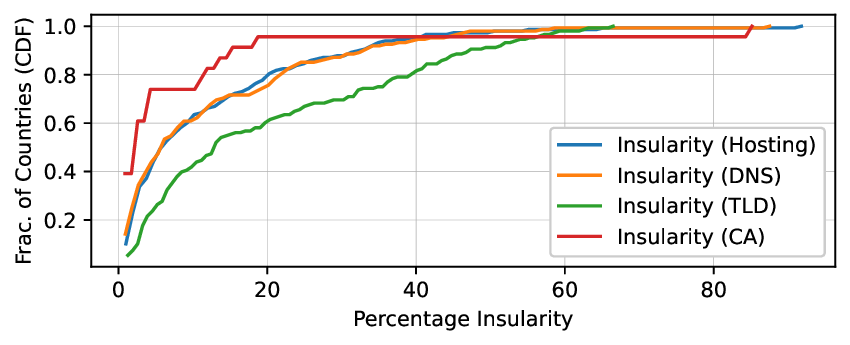}
    \vspace{-15pt}
    \caption{CDF of Insularity Across Layers---\textnormal{Countries tend to be more insular in their usage of TLDs compared other layers. Insularity in hosting and DNS track closely to each other. The small number of CAs and the relative domination of the large global CAs lead to a skewed distribution of insularity with the U.S. being the most insular.}}
    \label{fig:insularity_cdf}
\end{figure}
}

\newcommand{\insularityhostingbarplot}{
 \begin{figure*}[t]
    \centering
    \includegraphics[width=\linewidth]{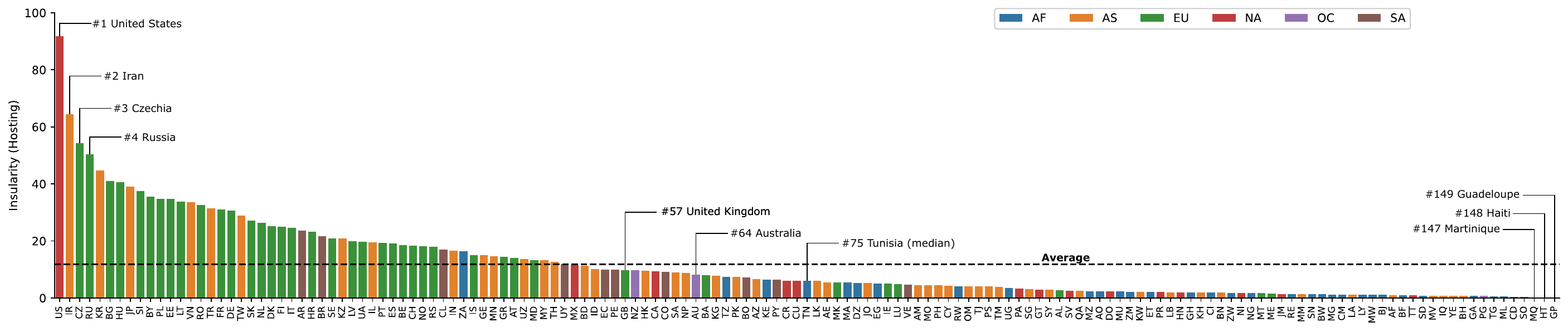}
    \caption{Hosting Provider Insularity---\textnormal{Countries sorted by \% of websites served by hosting providers based in the same country color coded by continent. The US is the most insular. While Iran is the second most insular it is also the least centralized because the top websites use a long tail of local providers. Countries in Europe tend to be more insular while countries in Africa do not show much insularity primarily due to a lack of equivalent providers in the country. }}
    \label{fig:insularity_barplot}
\end{figure*} 
}

\newcommand{\insularitydnsbarplot}{
 \begin{figure*}[t]
    \centering
    \includegraphics[width=\linewidth]{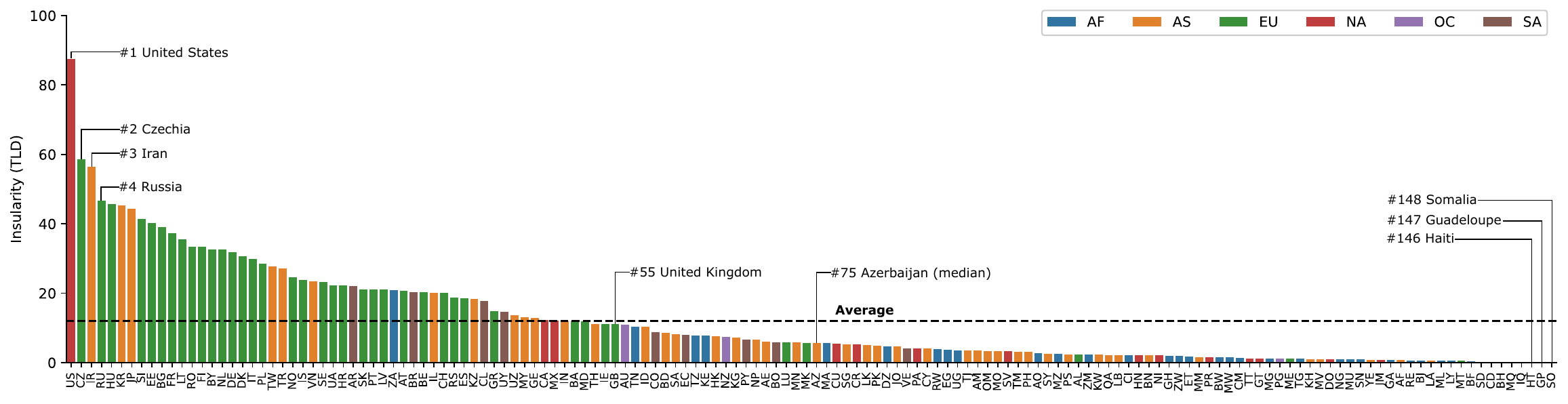}
    \caption{DNS Insularity by Country---\textnormal{Countries sorted by \% of websites served by DNS infrastructure providers in the same country color coded by continent. The DNS shows similar trends to that of hosting providers.}}
    \label{fig:insularity_dns_barplot}
\end{figure*} 
}

\newcommand{\insularitycabarplot}{
 \begin{figure}[h]
    \centering
    \includegraphics[width=\linewidth]{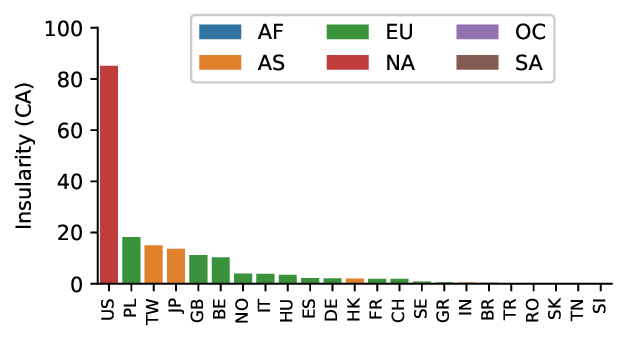}
    \caption{CA Insularity by Country---\textnormal{Countries sorted by \% of websites using CAs based in the same country, color coded by continent. Only 24 countries in our dataset show any use of a CA based in their own country. Unlike other layers with many regional providers, large global CAs based in the US dominate the ecosystem, with smaller CAs primarily used in a handful of countries for specific use cases.}}
    \label{fig:insularity_ca_barplot}
\end{figure} 
}

\newcommand{\insularitytldbarplot}{
 \begin{figure*}[t]
    \centering
    \includegraphics[width=\linewidth]{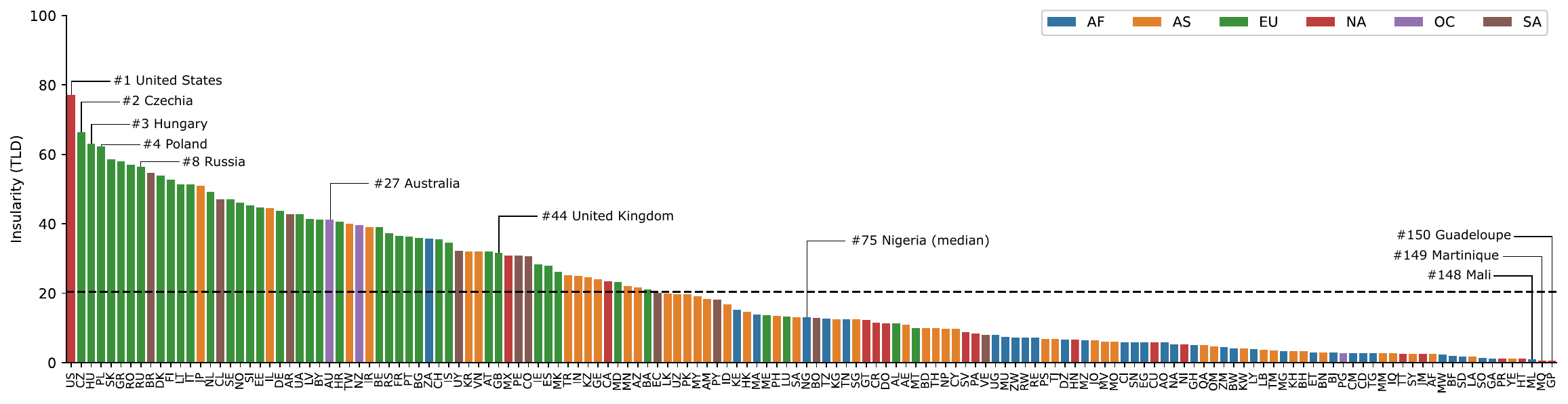}
    \caption{TLD Insularity by Country---\textnormal{Countries sorted by their \% use of their own local ccTLD. We consider the use of \dns{.com} to be insular to the US given the historical role of the US Government until recently in its operation~\cite{icann_usdoc_com}. Countries in Europe (especially Eastern Europe), Eastern Asia, and South America (especially Brazil) make heavy use of their local ccTLDs.}}
    \label{fig:insularity_tld_barplot}
\end{figure*} 
}

%% file: tables.tex
\newcommand{\dnsclasstable}{
\begin{table}[]
\small
\begin{tabular}{lrll}
\toprule
\textbf{Class} & \multicolumn{1}{c}{\textbf{Count}} & \textbf{Description} & \textbf{Example} \\
\midrule
XL-GP & 2 & Extra Large Global & Cloudflare \\
L-GP  &  10 & Large Global &  NSONE \\
L-GP (R) & 2 & Large Global (Regional) & OVH \\
M-GP &  17 & Medium Global &  DNSimple \\
S-GP &  78 & Small Global &  Sucuri \\
L-RP &  273 & Large Regional &  Alibaba \\
S-RP &  578 & Small Regional &  Scalaxy \\
XS-RP & 9,049 & Extra Small Regional & Forthnet \\
\bottomrule
\end{tabular}
    \caption{Classes of DNS Infrastructure Providers---\textnormal{As before, we categorize providers by usage and endemicity.}}
    \label{tab:dns-clusters}
\end{table}
}

\newcommand{\topfivetable}{
\begin{table*}[]
\small
\begin{subtable}{0.24\textwidth}
    \centering
    \begin{tabular}{rccr}
    \toprule
    \multicolumn{1}{c}{Rank} & \multicolumn{2}{c}{Country} & \multicolumn{1}{c}{\cs} \\ 
    \midrule
    \#1 & TH & AS & 0.3548 \\
    \#2 & ID & AS & 0.3258 \\
    \#3 & MM & AS & 0.2641 \\
    \#4 & LA & AS & 0.2526 \\
    \#5 & IQ & AS & 0.2490 \\
    \midrule
    \#146 & CZ & EU & 0.0561 \\
    \#147 & RU & EU & 0.0554 \\
    \#148 & SK & EU & 0.0497 \\
    \#149 & TM & AS & 0.0461 \\
    \#150 & IR & AS & 0.0411 \\
    \bottomrule
    \end{tabular}
    \caption{Hosting Providers
    } 
    \label{tab:centralized_top_countries_all}
\end{subtable}
\begin{subtable}{0.24\textwidth}
    \centering
    \begin{tabular}{rccr}
    \toprule
    \multicolumn{1}{c}{Rank} & \multicolumn{2}{c}{Country} & \multicolumn{1}{c}{\cs} \\ 
    \midrule
    \#1 & ID & AS & 0.3757 \\
    \#2 & TH & AS & 0.3374 \\
    \#3 & IQ & AS & 0.2730 \\
    \#4 & SY & AS & 0.2653 \\
    \#5 & LY & AF & 0.2548 \\
    \midrule
    \#146 & TM & AS & 0.0460 \\
    \#147 & FI & EU & 0.0459 \\
    \#148 & SK & EU & 0.0429 \\
    \#149 & HU & EU & 0.0404 \\
    \#150 & CZ & EU & 0.0391 \\
    \bottomrule
    \end{tabular}
    \caption{DNS
    } 
    \label{tab:centralized_top_countries_dns}
\end{subtable}
\begin{subtable}{0.24\textwidth}
    \centering
    \begin{tabular}{rccr}
    \toprule
    \multicolumn{1}{c}{Rank} & \multicolumn{2}{c}{Country} & \multicolumn{1}{c}{\cs} \\ 
    \midrule
    \#1 & SK & EU & 0.3304 \\
    \#2 & CZ & EU & 0.3268 \\
    \#3 & EE & EU & 0.2811 \\
    \#4 & IR & AS & 0.2807 \\
    \#5 & SI & EU & 0.2623 \\
    \midrule
    \#146 & KR & AS & 0.1631 \\
    \#147 & CO & SA & 0.1618 \\
    \#148 & VN & AS & 0.1599 \\
    \#149 & JP & AS & 0.1499 \\
    \#150 & TW & AS & 0.1308 \\
    \bottomrule
    \end{tabular}
    \caption{CAs
    } 
    \label{tab:centralized_top_countries_cas}
\end{subtable}
\begin{subtable}{0.24\textwidth}
    \centering
    \begin{tabular}{rccr}
    \toprule
    \multicolumn{1}{c}{Rank} & \multicolumn{2}{c}{Country} & \multicolumn{1}{c}{\cs} \\ 
    \midrule
    \#1 & US & NA & 0.5853 \\
    \#2 & PR & NA & 0.5358 \\
    \#3 & TT & NA & 0.4821 \\
    \#4 & JM & NA & 0.4771 \\
    \#5 & CZ & EU & 0.4656 \\
    \midrule
    \#146 & KZ & AS & 0.1629 \\
    \#147 & UZ & AS & 0.1569 \\
    \#148 & TJ & AS & 0.1526 \\
    \#149 & MD & EU & 0.1475 \\
    \#150 & KG & AS & 0.1468 \\
    \bottomrule
    \end{tabular}
    \caption{TLDs
    } 
    \label{tab:centralized_top_countries_tld}
\end{subtable}
\caption{Top 5 most centralized and least centralized countries for each layer under analysis, based on the top 10K sites in each country. Note the commonalities between hosting and DNS centralization. Iran, Slovakia, and Czechia were the least centralized from the perspective of hosting providers but one of the most centralized from the perspective of CAs.}
\end{table*}
}

%% file: intro.tex
\section{Introduction}

Following the widespread adoption of cloud hosting and content distribution networks (CDNs), Internet centralization has become a hot topic of debate in the Internet standards and measurement communities over the past decade. On the one hand, many providers argue that their services improve performance, resilience, and security for their customers and, in turn, their users~\cite{cfnet, fastlynet}. On the other hand, members of the networking community have expressed concern over the increasingly centralized nature of the Internet, arguing that the Internet's original decentralization was one of the foremost reasons for its success and warning that centralization can limit innovation, prevent competition, reduce availability, and pose privacy concerns~\cite{rfc_decentralization, imc20_dependencies_mirai, centrality_dns}.



Despite increased attention to the topic, most analyses of centralization rely on informal, intuitive definitions of centralization and descriptive statistical measures~\cite{toit_consolidation, noms22_hosting_consolidation, imc20_clouding_up_internet, centrality_dns, imc20_dependencies_mirai, each_at_its_own_pace}. These approaches have unquestionably shed light on Internet behavior, but, without a rigorous metric for quantifying centralization, we are precluded from performing much-needed in-depth analysis. In this work, we formalize a statistical measure of centralization, which we use to understand how the web has centralized globally.

Building on our community's informal agreement of centralization as \emph{the
concentration of an Internet function on a small number of providers}, we start by formulating a set of requirements for a quantitative measure of Internet centralization. We then formalize centralization as \textit{the statistical distance of
an observed distribution from a fully decentralized reference distribution}, which we quantify using Wasserstein distance, a popular integral probability metric.

With this grounding, we measure the centralization of four aspects of the web---hosting/content delivery, authoritative DNS infrastructure, top-level domains, and certificate authorities---in 150~countries. Consistent with prior work, we observe the key role that the largest global providers like Cloudflare and Amazon play in Internet centralization. However, we also find that countries vary dramatically in their centralization. For example, the most centralized country (Thailand) relies on a single hosting provider for 60\% of its top websites, while the least centralized (Iran) has no provider serving more than 14\% of its popular sites.

Our analysis exposes that centralization is intrinsically linked with two other notions, insularity and regionalization. Many countries' dependencies are sociopolitical, which shapes the set of providers they rely on. To better understand this relationship, we classify providers along two axes: scale and global reach. We explore the contributions that global and regional providers make to centralization: although reliance on regional providers usually correlates with a more diffuse and thus less centralized provider ecosystem, we also find that regional providers in many countries hold significant sway, rivaling the market share of the major global players. Indeed, beyond the very largest providers like Cloudflare and Amazon, large global providers often have less influence on the centralization of a country than large regional providers that have gone unnoticed by centralization studies focusing only on globally aggregated data.

Our measures additionally shed light on provider dependencies within and between countries, a form of consolidation that affects users. 
We highlight regional patterns that paint a richer picture of provider dependence and insularity than we can through centralization alone. For instance, the Commonwealth of Independent States (CIS) countries (formed following the dissolution of Soviet Union) exhibit comparatively low centralization, but depend highly on Russian providers. These patterns suggest possible political, historical, and linguistic undercurrents of provider dependence. In addition, the regional patterns we observe between layers of website infrastructure enable us to hypothesize about forces of influence driving centralization across multiple layers. For example, many countries are more insular in their choice of TLD given the limited technical implications of TLD choice. On the other extreme, certificate authority (CA) centralization is far more extreme than other layers due to popular web browsers trusting only a handful of CAs, nearly all of which are located in the United States.

We conclude with observations across layers and recommendations for the Internet community. Beyond the analysis in our work, we hope that by rigorously defining a statistical measure of centralization, we can enable further study on this critical topic.

%% file: background.tex
\section{Related Work}
\label{sec:background}

There is a growing body of literature documenting centralization of different layers of the web, including cloud and hosting platforms~\cite{noms16_apples_oranges, toit_consolidation,
noms22_hosting_consolidation, arxiv_feamster_marshini_2024, holz2020tracking, ccr_web_still_small}, DNS~\cite{centrality_dns, internet_entropy,
imc20_clouding_up_internet}, certificate authorities
(CAs)~\cite{imc20_dependencies_mirai,each_at_its_own_pace}, and third-party web
resources~\cite{www20_third_party, arxiv_third_party}. For example, Doan et~al.~\cite{toit_consolidation} study the web content served by hosting entities (\eg,~fonts, and ads), finding that hosting services' increased dominance allows them to contribute to popularizing new Internet standards. Hoang et~al.~\cite{ccr_web_still_small} find pronounced domain co-hosting at large providers. Several other studies highlight provider consolidation and single points of failure as a concern for web resiliency~\cite{acsac_leanonme, imc20_dependencies_mirai, internet_entropy, arxiv_feamster_marshini_2024}. 

While most centralization research has focused on global concentration on providers, a number of studies have hinted at country-level differences~\cite{helles2020infrastructures, each_at_its_own_pace, kende2015promoting}. Kumar et~al.~\cite{each_at_its_own_pace} analyze the Alexa Top~500 sites in 50~countries, finding that centralization has increased over time and is correlated with measures of economic development. Zembruzki et~al.~\cite{noms22_hosting_consolidation} study hosting centralization for websites in 19~TLD zone files, finding evidence of geographic and language patterns in hosting provider dependence. Another set of work has examined web infrastructure within specific regions. For instance,
Jonker et~al.~\cite{imc22_whereru} describe Russian domain infrastructure.
Fanou et~al.~\cite{fanou2016africa, fanou2018africa} describe web infrastructure in Africa, including regional network infrastructure of major CDNs and trends of hosting abroad.
Helles et~al.~\cite{helles2020infrastructures} cluster top-150 EU websites with similar third-party dependencies and show regional variation in third-party dependence.
Looking more globally, Fan et~al.~\cite{fan2015assessing} measure how CDNs map users to front-end clusters and where content is served. Our work is complementary, analyzing how centralization differs across regions and how unique socioeconomic patterns shape centralization.

Prior studies on web centralization have primarily relied on summary statistics, ranging from collections of raw
numbers~\cite{toit_consolidation, noms22_hosting_consolidation,
imc20_clouding_up_internet, noms16_apples_oranges, ccr_web_still_small, acsac_leanonme, holz2020tracking}, to top-k
thresholds~\cite{centrality_dns, imc20_dependencies_mirai, toit_consolidation,
each_at_its_own_pace, noms16_apples_oranges, arxiv_feamster_marshini_2024}, to custom defined metrics that focus on a handful of
providers~\cite{toit_consolidation}. While these works have unquestionably shed light on centralization, our work shows how a more quantitative, statistical approach can help us to uncover nuances in how centralization is occurring. Most closely related to us, Huston uses Herfindahl-Hirschman Index (HHI) in a blog post looking at the use of recursive resolvers; Bates et~al.~\cite{internet_entropy} also use HHI in a National Bureau of Economic Research whitepaper looking at redundancy of DNS servers. 

Last, there is also a collection of work that looks at centralization beyond the web. For example, Moura et~al.~\cite{imc20_clouding_up_internet} analyze DNS queries seen at authoritative DNS servers to investigate centralization on cloud providers. While some works study centralization directly, others examine distributions of providers in service of complementary questions. 
Clark and Claffy~\cite{clark2021trust} discuss topological, rather than geographical, regionalization in Internet routing. Dell'Amico et~al.~\cite{acsac_leanonme} use graph statistics crafted to their use case of measuring web resiliency. While there has also been a significant body of work studying concentration trends in network topology (e.g.,~\cite{gill2008flattening, bottger2019ixp}), here we address settings in which the data is not necessarily a graph (e.g., a count of how many websites rely upon certain providers to function).


While this large body of prior work has effectively called attention to
the issue of centralization, our work argues that to move the discussion forward, we need to establish a common framework to quantify and report on centralization as experienced by users across different countries.

%% file: centralization.tex
\section{Quantifying Centralization and Regionalization}
\label{sec:centralization}

In this section, we formalize a statistical definition of centralization, as well as introduce two additional measures, endemicity and usage, to characterize the reach and popularity of providers. We then describe how we use these metrics to measure the centralization and regionalization of the web.

\subsection{Defining Centralization}


\emdscoreexample

While the the Internet community has not formalized a statistical definition of
centralization, there is informal agreement that centralization is \emph{the
concentration of an Internet function on a small number of providers}.
%
While intuitive, this definition is difficult to quantify: ``concentration'' and
``small'' are ill defined, limiting researchers to 
self-evident cases and hindering principled analyses like cross-country
comparisons. To design a more quantitative metric, we translate
the above definition into a set of requirements. Our metric should:

\vspace{-3pt}
\begin{enumerate}
    \item account for both the number of providers and the distribution of the the Internet function across those providers in a single metric.
    \item accommodate comparing highly skewed distributions with few dominant providers and a long tail of others. 
    \item facilitate fair comparisons across scenarios (\eg, across countries), independent of the specific providers underlying the distribution. 
    \item match human intuition of centralization in the context of Internet functionality.
\end{enumerate}
\vspace{-3pt}


\noindent In this work, we formalize centralization as \textit{the distance of
an observed distribution from a fully decentralized reference distribution}.
Namely, we compare observed provider distributions against a
hypothetical uniform distribution where every Internet resource has a unique
provider.  Our decentralized distribution does not express an ``ideal'' or even
a possible state of the world, but rather serves as a reference relative to which
all other distributions can be compared; the observed distribution with the
largest distance from the fully-decentralized distribution is the most
centralized.

This formulation allows us to consider a wide range of statistical distance measures, including $f$-divergence functions (\eg,
Kullback--Leibler (KL) divergence, Jensen--Shannon divergence, Hellinger
distance, and total variation distance) and integral probability metrics (\eg, Wasserstein distance, Dudley
metric, and maximum mean discrepancy). 
While $f$-divergence functions like KL-divergence are popular for measuring distribution dissimilarity, we find them unsuitable for our particular task. The $f$-divergence class works well only when comparing two distributions that are largely overlapping (the $f$-divergence between any two fully disjoint distributions is constant). In our case, we are comparing two fundamentally different distributions: a real-world observed distribution that is heavily skewed towards a few large providers and a hypothetical uniform distribution where every service has its own provider. We note that we cannot simply compare observed provider distributions directly against each other since there would not be a clear directionality as to which distribution is more centralized; instead, we compare all observed distributions against a reference distribution of absolute zero centralization. 
Ultimately, we decide to use Wasserstein's distance, one of the---if not
the---most commonly used integral probability metrics~\cite{cai2022distances}
that captures the distance between distributions even when they do not
significantly overlap.

\vspace{-0.03in}
\subsection{Quantifying Centralization} 
\label{sec:emd_cs}
%
We use Wasserstein distance---also commonly known as \emph{Earth Mover's
Distance} (EMD) in the computer vision
community~\cite{cv_emd,ijcv_emd_cv}---to measure centralization. The metric
earns this name because it effectively considers two distributions as a mass
of earth spread out in space and computes the minimum amount of ``work'' needed
to transform the first distribution into the second, where a unit of work
corresponds to transporting a unit of earth by a unit of ground distance
according to a customizable ground distance function. The total work is the
product of the mass of earth moved and the distance it was moved.  

We apply EMD to our use case by measuring the work that would be required to transform an observed distribution of providers $A = (a_1,
\ldots, a_n)$ into the completely decentralized reference distribution $R = (r_1, \ldots, r_m)$ where every
website uses a unique provider.  Thus $r_j = 1 \ \forall j$, and $m=C$ where $C = \sum_i a_i$ is
the total number of websites considered.  
We define our ground distance metric $d_{ij}$ between $a_i \in A$ and $r_j \in R$ as the difference between $a_i$ and $r_j$, normalized by the total number of websites.
Note, then, that websites using more popular providers must travel a greater distance toward a fully decentralized distribution compared with websites using less popular providers. Intuitively, the distribution that is the most centralized is the one that would require the greatest work to become fully decentralized and thus has the largest value for our centralization score (Figure~\ref{fig:centralization_example}). 





Writing out this definition more formally yields the following expression for our \CS (\cs):
$$\textnormal{\cs} = \sum_{i=0}^n\left(\frac{a_i}{C}\right)^2 - \frac{1}{C}$$
Note that we can read the $\frac{a_i}{C}$ term as the fraction of websites that provider $i$ hosts, and that the upper bound on \cs is $1-\frac{1}{C}$, which approaches 1 as a larger $C$ is chosen.  Readers familiar with the Herfindahl–Hirschman Index (HHI), a metric used to measure competition between market entities in antitrust law~\cite{doj_hhi,doj_hhi_merger} will recognize $\sum_{i=0}^n\left(\frac{a_i}{C}\right)^2$ as the formula for HHI applied to our setting. Thus, HHI is effectively a special case instantiation of EMD up to a constant. Further details are provided in Appendix~\ref{sec:emd:math}. This instantiation of EMD fulfills our aforementioned goals for a centralization metric:

\begin{enumerate}
    \item The score accounts for both the size of providers (the larger the provider, the more it contributes to \cs) and the number of providers (the longer the tail, the more providers contribute small $d_{ij}$ terms to the sum).
    \item Our metric makes no assumptions about the similarity or overlap between the distributions we are comparing, and provides a meaningful measure of distance even for highly skewed distributions.
    \item As long as $C$ is held constant throughout an analysis, the metric can fairly compare against multiple scenarios: it is based only upon the shape of the curve, not on the providers comprising it.
    \item The notion of quantifying the ``work'' required to flatten the data into a fully decentralized distribution is appealing in the context of centralization: the more concentrated the the underlying distribution, the more ``work'' and the greater the \cs~ value. The largest providers are weighted most heavily in the metric: a provider's contribution to \cs~ grows quadratically with its market share among the set of websites considered, which empirically we find matches our intuition of which providers most contribute to centralization.
\end{enumerate}

\noindent While we do not specify a cutoff for how large of 
\cs~ corresponds to ``centralized,'' we provide a set of example curves to aid intuitive interpretation of \cs~ values in Figure~\ref{fig:interpretation_guide}.
%
Since \cs~ is closely related to HHI, the HHI interpretation guidelines used by the U.S. Department of Justice
for antitrust analysis can also provide context of how other fields think about concentration: `competitive' ($<$0.10),
`moderately concentrated' (0.10--0.18), and `highly concentrated'
($>$0.18)~\cite{doj_hhi,doj_hhi_merger}. 


\interpguide

Last, we note that EMD is customizable and provides a framework for future
work to consider other aspects of Internet centralization. While we choose a
specific reference distribution $R$ and ground distance function $d_{ij}$ that
suit our goals, future work may want to explore other parameters. For
example, 
a study looking at how countries rely on specific providers may wish to redefine
$d_{ij}$ and compare countries' distributions pairwise rather than using a
reference distribution. Another natural extension would be to assign an
unequally weighted ``mass'' to each website (\eg, based on traffic), rather
than weighting all sites equally.

\subsection{Quantifying Regionalization}
\label{sec:endemicity}

To add explanatory depth of how the Internet is centralized, we need a language
to describe the reach of the providers we encounter. Providers vary both in
their overall usage and in their geographic usage distribution.
To illustrate these differences, we consider each provider's \textit{usage curve} similar to Ruth \etal~\cite{imc22_ruth_crux_world_wide_view}: we compute the percentage of popular websites in each country that use the provider, sort countries by decreasing usage, and plot the resulting list of usage values. We write these values as a nonincreasing sequence $(u_1, u_2, \ldots, u_n)$ where $n=150$ is the total number of countries considered. Figure~\ref{fig:provider_usage_curve} shows two examples from our hosting provider data (from Section~\ref{sec:country_hosting}). The first is of a large globally popular provider like Cloudflare, which is used by many websites across many countries; the second is of Beget LLC, a Russian provider mainly used in Commonwealth of Independent States (CIS) countries formed following the dissolution of the Soviet Union.  
%

\usagecurve

\vspace{2pt}
\paragraph{Usage} We define \textit{usage} ($U$) to be the area under the usage curve: $U=\sum_{i=1}^{n}u_i$. This metric captures total usage across the countries in the dataset. In Figure~\ref{fig:provider_usage_curve}, the total usage of the example global provider is {larger} than the regional one.

\vspace{2pt}
\paragraph{Endemicity} Second, we define \textit{endemicity} ($\mathcal{E}$) to be the area between the usage curve and the flat line starting at the usage curve's maximum value: $\mathcal{E} = \sum_{i=1}^{n} (u_1-u_i)$. This captures the deviation from absolute global consistency in usage (i.e., the ``flatness'' of the usage curve), with a priority on capturing unusual popularity in a country rather than unusual unpopularity. We adopt this metric from Ruth et~al.~\cite{imc22_ruth_crux_world_wide_view}, with one modification: we normalize by provider size by computing the \textit{endemicity ratio} as $\mathcal{E}_R = \frac{\mathcal{E}}{U+\mathcal{E}}$. Without this modification, the range of possible endemicity values is a function of the provider's maximum percent use in any country. Though endemicity ratio is not the only method of correcting for this~\cite{imc22_ruth_crux_world_wide_view}, we find it to be the most natural for our use case. Values of $\mathcal{E}_R$ range from $0$ to $1$, with smaller values indicating more global reach and larger values indicating regional concentration. In Figure~\ref{fig:provider_usage_curve}, the example regional provider is \textit{more endemic} than the global one.
In summary, the area under the usage curve tells us how ``large'' the provider
is globally, while the endemicity ratio captures how ``regional'' versus ``global'' the provider is.

\subsection{Data Collection}
\label{sec:datasets}

Equipped with our definitions of centralization, usage, and endemicity, we
analyze the centralization and regionalization of several layers of the web: (1)
hosting providers (Section~\ref{sec:country_hosting}), DNS providers
(Section~\ref{sec:country_dns}), TLDs (Section~\ref{sec:country_tld}), and Certificate Authorities
(Section~\ref{sec:country_ca}), across
150~countries (Appendix~\ref{app:country_references}). Here, we describe how we collect and enrich our data.

\vspace{2pt}
\paragraph{Popular Websites}
Our analysis is based on the public Chrome User Experience Report (CrUX)~\cite{crux}, which lists the most popular websites per country, grouped in rank-magnitude buckets; prior work has shown CrUX to most accurately capture popular websites~\cite{imc22_crux_eval_top_lists}. Due to differing traffic volumes per country and Google's safeguards to protect user privacy, per-country popularity lists differ in length: smaller countries and countries with lower Chrome adoption have fewer websites in the dataset. To facilitate comparisons across countries, we analyze the top 10K~websites in each of 150~countries whose top lists are at least that long. This covers a substantial fraction (63.3\%) of countries and a substantial fraction of traffic on the internet: the top 10K websites cover about 70--80\% of web traffic globally~\cite{imc22_ruth_crux_world_wide_view}. 

\vspace{2pt}
\paragraph{Collecting DNS and Certificate Data}
Building on the CrUX data, we perform active measurements for each website and
annotate this dataset with additional third party data. We resolve the
588K~domains using ZDNS~\cite{izhikevich2022zdns} during May 2023, geolocate IPs
with NetAcuity~\cite{netacuity_geolocation} and add origin AS using
pfx2as~\cite{caida_pfx2as}, AS WHOIS organization and country using CAIDA's AS
to Organization tool~\cite{as2org}, and anycast configuration using
\dns{bgp.tools} anycast dataset~\cite{bgptools_anycast}. Finally, we use
ZGrab2~\cite{durumeric2015search} to attempt a TLS handshake with the website on each resolved
IPv4 address; following each handshake, we parse the leaf certificate and label
CA ownership using the Common CA Database (CCADB)~\cite{ccadb-country} per Ma
et~al.~\cite{ma2021s}. These measurements are performed from a vantage point in a
large university network. Using this enriched toplist data, we calculate hosting provider
centralization using the AS Organization of the IP address serving the
content, the DNS infrastructure centralization using the AS Organization of the
nameserver IP address, the CA centralization using the CA Owner for the certificate
served at the hosting IP.

\vspace{2pt}
\paragraph{Interpreting Statistics} We apply our numerical definitions of centralization (\cs), usage ($U$), and endemicity ratio ($\mathcal{E}_R$) as defined in Sections~\ref{sec:emd_cs} and~\ref{sec:endemicity} to our data on hosting, DNS, CA, and TLD use. In addition, we use Pearson's correlation coefficient $\rho$ as needed to compute correlation between rank-ordered sequences. We follow these guidelines for interpreting correlation coefficients: $<$0.30 is poor, 0.30--0.60 is fair, 0.60--0.80 is moderate, and $>$0.80 is strong~\cite{correlation_guide}.

\vspace{2pt}
\paragraph{Limitations} Our analysis relies on Chrome website popularity data, which is compiled from Chrome telemetry data for users who opted in to data sharing; it may exhibit bias due to excluding incognito browsing, and it may be less representative of countries with lower rates of Chrome adoption. Nevertheless, prior work has shown that it is the most accurate data available~\cite{imc22_crux_eval_top_lists}.
For our active measurements, our method for determining hosting providers relies on the AS Organization, and as such we only see the last leg of content delivery. As this method cannot distinguish delivery from the underlying hosting provider (e.g., in the case of CDNs), we conflate the two and refer to both as ``hosting'' in this work. 
Finally, we perform all active measurements from a single vantage point; we encourage future replication studies from other vantage points.

%% file: country_provider.tex
\section{Hosting Providers}
\label{sec:country_hosting}




\cdfcountryexample


In this section, we explore how centralized and regionalized the web is in
terms of the hosting providers that serve popular websites. 
Note, we consider the provider that serves content for each website. Since many
websites use CDN or DDoS protection providers like Cloudflare, we can only
determine the forward-facing provider and not the backend hosting provider. In
this paper, we refer to the AS Organization serving the website 
as the hosting provider.

%
%
%


\csproviderbarplot

\subsection{Centralization of Hosting}
\label{subsec:hosting_overview}

Consistent with prior work, we find evidence of concentration of hosting on a small number of providers: 90\% of websites are hosted by fewer than 206~providers in every country. However, as can be seen in Figure~\ref{fig:centralization_barplot}, there is
significant variance in the degree to which countries have centralized. In the
most extreme (\ie, most centralized) case, 60\% of websites in Thailand 
(\cs = 0.3548) are served by a single provider, Cloudflare.  
At the other extreme, Iran is the least centralized country with
the top provider accounting for only 14\% of websites and 90\% of websites
distributed across 80~providers (\cs = 0.0411).  The United States is the
median country of the dataset ($\tilde{\mathscr{S}}$ = 0.1358). We provide a full list  in Appendix~\ref{sec:cs_scores}. 


While it is difficult to predict how centralized an individual
country will be, there are subregional patterns (Figure~\ref{fig:cs_boxplots_regions}).  Southeast Asia is the most centralized subregion ($\bar{\mathscr{S}}$
= 0.2403) and includes the four most
centralized countries: Thailand, Indonesia, Myanmar, and Laos. In contrast, Central Asia is the least centralized 
($\bar{\mathscr{S}}$ = 0.0788), with Turkmenistan as the second-least
centralized country.
European countries are on the whole consistently less centralized,
with Slovakia, Russia, and Czechia in Eastern Europe being the least centralized
countries in the region (\csregion{EU} = 0.0994, \csregion{E-EU} =
0.0803). 
While Africa, as a whole, does not tend toward any extreme
(\csavg = 0.1553), countries in Northern Africa (Libya)
and Eastern Africa (Somalia) are more centralized
than the global average (\csavg = 0.1429, var = 0.003). As
we will discuss in Section~\ref{sec:hosting:regionalization:continent}, this is largely due to using 
providers on other continents.

The total providers in a country has little effect on how
centralized the country is: Thailand (most centralized) and Iran (least
centralized) have the second (328) and sixth (444) fewest providers while the U.S. (median
centralization) has the fourth most providers (834). 
Rather, centralization is most heavily affected by
the breakdown of market share amongst the top providers in that country. 
%
%
When we break down the distribution amongst the top ten providers in
each country, we see significant variation. The top provider in Thailand
operates 60\% of top websites, whereas the top provider in the U.S. operates only
29\%, and in Iran 14\% of websites. After the top 10--100 providers,
countries have varying long tails of providers. For example, providers with fewer than 100~websites in our dataset host 17\% of Iran's top sites but only 8\% of Thai websites. 


Anecdotally, we note that the most centralized countries tend to rely on
extremely large providers like Amazon and Cloudflare, whereas the least centralized
countries rely more greatly on regional providers. We investigate how different
types of providers (\eg, large global vs.\ small regional providers) affect
country centralization in the next section.

\subsection{Classes of Providers}
\label{sec:hosting:providers:centralization}
\label{subsec:hosting_categories}

Countries vary not only in their concentration on popular providers, but in
the types of providers they use. For example, while seven of the top ten
providers in Thailand are large global providers, seven of the top ten
providers in Iran are local to the country. 
%
%
To understand the types of hosting providers that countries depend on and how
this affects centralization, we build classes of hosting providers by calculating the usage and endemicity ratio for each provider
%
%
and then apply min-max scaling and cluster using affinity propagation.
This results in 305~clusters (Figure~\ref{fig:categorizing_providers_clusters}), which we manually examine to identify 8~classes of providers (Table~\ref{tab:hosting-clusters}).



\hostingclusters

\begin{table}[]
\small
\begin{tabular}{lrll}
\toprule
\textbf{Class} & \multicolumn{1}{c}{\textbf{Providers}} & \textbf{Description} & \textbf{Example} \\
\midrule
XL-GP & 2 & Extra Large Global & Cloudflare \\
L-GP & 6 & Large Global &  Akamai \\
L-GP (R) & 2 & Large Global (Regional) & OVH \\
M-GP &  22 & Medium Global &  Incapsula \\
S-GP &  73 & Small Global &  Wix \\
L-RP &  174 & Large Regional &  Alibaba \\
S-RP &  587 & Small Regional &  Loopia \\
XS-RP & 11,548 & Extra Small Regional & Forthnet \\
\bottomrule
\end{tabular}
    \caption{Classes of Hosting Providers---\textnormal{We categorize providers based on how many top websites they serve (usage) and their usage consistency across countries (endemicity).}
    }
    \label{tab:hosting-clusters}
    \vspace{-12pt}
\end{table}


\providerstackedbarplot


One of the two XL-GP providers is the largest provider in every country. With the
exception of Japan, which relies most on Amazon, Cloudflare is the top provider
in every country. As can be seen in Figure~\ref{fig:hosting_provider_stacked},
the centralization of each country is strongly correlated with the dominance of
these two XL-GP providers (Pearson Correlation Coefficient $\rho$=0.90, p
$\ll$ 0.05). Perhaps surprisingly, the use of large global providers (L-GP) beyond
Amazon and Cloudflare (\eg, Google, Akamai, and Microsoft) has only a poor
correlation with centralization ($\rho$=0.19, p $<$ 0.05). This class of large
global providers serve 11--41\% of websites between countries, but never sees outsized adoption in any country; rather, websites are
split across a number of large global providers.

Beyond the global providers, countries' use of regional providers varies
substantially, from 12\% (Trinidad and Tobago, Caribbean) to 68\% (Iran). Among
the least centralized countries, a cluster of regional providers overshadows the
global providers, as shown in Figure~\ref{fig:hosting_provider_stacked}.
Generally, regional providers create diffuseness in the provider ecosystem:
indeed, the use of large regional providers is moderately correlated with lower
centralization in countries ($\rho$=$-$0.72, p $\ll$ 0.05). However, in a few
cases, there is a single dominant large regional provider, such as
SuperHosting.BG in Bulgaria (22\%, \cs = 0.1188) and UAB in Lithuania (22\%, \cs
= 0.1286), which never outrank Cloudflare but come a close second, 
contributing meaningfully to centralization in one country despite not
being significant players globally. 

\subsection{Regionalization and Insularity}
\label{sec:hosting:regionalization}




%
Most regional providers achieve substantial market share in only the country in which they are headquartered. However, at times, provider dependence follows other geopolitical patterns. In this section, we look at the insularity of countries and other countries they depend on.

%

\subsubsection{Insularity}

We define insularity as the fraction of websites in a country that
are hosted by providers headquartered in the same country. Statistically, the U.S. is most insular (92.1\%) because global providers are frequently American. This reliance also means that most countries are not very insular: U.S. providers host the largest share of sites in all but five countries: Iran, Czechia, Russia, Hungary, and Belarus.
Iran (\cs = 0.0411) with $64.8\%$, Czechia (\cs = 0.0561) with
$54.5\%$, Russia (\cs = 0.0554) with $51.1\%$ of websites hosted by local providers are the three most insular countries in our dataset.
The other two countries with little reliance on the U.S., Turkmenistan and Slovakia, are not particularly insular: only 4\%~of websites in Turkmenistan use in-country providers; rather they rely heavily on Russian providers ($33\%$). In, Slovakia, 25.7\% of sites use Czech providers.

More broadly, East Asia (South Korea and Japan) and Europe 
tend towards being more insular. Countries in Africa have low
insularity (average of 3\%), indicating dependence on external providers, which we explore in Section~\ref{sec:hosting:regionalization:continent}.
Insularity and centralization are different measures---a highly insular nation need not be highly centralized or decentralized. However, given that countries that do not rely on a single large global provider tend to have websites split across a handful of providers, {higher insularity is moderately correlated with lower
centralization} ($\rho$=-0.61, p $\ll$ 0.05).





\orggeoregion


\subsubsection{Regional Patterns}
\label{sec:hosting:regionalization:continent}

Grouping providers by the continent that they are headquartered in (Figure~\ref{fig:hosting_region_org_continent}), 
we find that European countries tend to rely on other European providers, with the exception of Northern Europe (\eg, UK, Sweden), which primarily relies on large global
providers. While Asia broadly uses global providers, Eastern Asia (\eg, Japan, South Korea) uses regional Asian providers and Central Asia (\eg, Turkmenistan and other former USSR countries) use
Russian providers (Section~\ref{sec:hosting:regionalization:clusters}). Top websites in
Africa are primarily hosted by American and European providers.

We stress that the use of providers headquartered in one country does not necessarily mean that the hosted content
is not geographically proximate to users. Breaking down top websites by the continent
in which the IP address is geolocated,
we see that many websites using North American providers
(Figure~\ref{fig:ip_geo_region_continent}) are located in the same region as the website. 
While not an important factor in hosting, anycast plays an important role in the DNS ecosystem. In fact, we find that some European countries 
use large global DNS providers while relying on
large regional providers for hosting (Section~\ref{sec:country_dns}).
%
%
Majority of websites in African countries are hosted in North America and Europe. It is unclear whether this is due to heavy use of American and European sites or if African sites use American or European hosting services.

%

\subsubsection{Regional Case Studies}
\label{sec:hosting:regionalization:clusters}

Looking beyond the evident dominance of large providers in the US, we see regional dependencies emerge. We manually analyze cases where countries rely disproportionately on select countries, and identify the following patterns:


%


\paragraph{Russia} Russian providers are heavily used by Commonwealth of Independent States (CIS): Turkmenistan (33\%), Tajikistan (23\%), Kyrgyzstan (22\%), Kazakhstan
(21\%) and Belarus (18\%); 
 not all post-Soviet states depend heavily on Russia: Ukraine (2\%), Lithuania (3\%), and Estonia (5\%).

\paragraph{France} France is the second most relied on popular country after the US, largely due to global reliance on OVH\@. Beyond OVH, eight French large regional providers (\eg, Online S.A.S) are extensively used in French administrative regions: Réunion Island (36\%), Guadeloupe (34\%), and Martinique (35\%); 
several African countries that were formerly French colonies, such as Burkina Faso (21\%), Côte d'Ivoire (18\%),  and Mali (18\%), also rely heavily on French hosting providers. 

\paragraph{Czechia} 26\% of the Slovakian top websites are hosted by Czech providers. Czechia, which is quite insular,
does not rely heavily on Slovakia.\footnote{Note that Czechoslovakia was a single country until 1992.}

\paragraph{Germany} The large German provider Hetzner is used globally (2\% of all sites). There are
also 7 large regional providers that see use in Austria (3\%) as well, which is
consistent with prior work \cite{noms22_hosting_consolidation} that attributes
this to the fact that German is the predominant language in both countries.

\paragraph{Iran} More than 20\% of Afghan top sites are hosted by Iranian providers. This may be due to shared language (Persian); we note that 31.4\% of the websites in
Afghanistan's top list are in Persian, of which 60.8\% are hosted in
Iran.\footnote{We detect the language of the website using
LangDetect~\cite{langdetect}.}


\subsection{Summary}

European countries are consistently less centralized, while Southeast Asian countries are more centralized. Cloudflare usage frequently drives the centralization in a country. Conversely, we find that use of regional providers is correlated with lower centralization in the country.
%






%% file: country_dns.tex
\section{DNS Infrastructure}
\label{sec:country_dns}




DNS centralization is broadly similar to hosting
centralization,\footnote{DNS Infra. centralization score by country: Appendix
\ref{subappendix:continent_rest}, Figure \ref{fig:centralization_dns_barplot}.}
in large part because most websites use the same provider for hosting and DNS.
Indeed, Cloudflare content serving is predicated on using their DNS service in most cases~\cite{cloudflare_manage}.
In the most centralized case, 65\% of the websites (up from 57\% in
hosting) in Indonesia (\cs = 0.3757) are served by
Cloudflare.\footnote{DNS provider type usage in all countries: Appendix \ref{subappendix:stacked_rest}, Figure \ref{fig:hosting_dns_stacked}.} Thailand, which was the most centralized country from the
perspective of hosting providers is a close second with 62\% of the websites (up
from 60\%) served by Cloudflare. In Czechia (\cs = 0.0391), the
least centralized country, Cloudflare usage remains the same as its use as hosting provider (17\%).
In contrast, Iran, drops to being the tenth least centralized country
given an increased use of Cloudflare. 
%
%
In addition to Cloudflare, some countries that depend on global
providers for hosting shift to large global providers that provide managed DNS services
such as NSONE, and Neustar UltraDNS (Section~\ref{sec:country_dns:classes}) and countries that depend on small regional providers for hosting often shift to larger regional providers for DNS.
%
Even with these underlying shifts, the broader regional trends remain largely
unchanged from hosting providers.

\subsection{Classes of Providers}
\label{sec:country_dns:classes}
Similar to hosting, we classify providers based
on their usage and endemicity ratio metrics
(Table~\ref{tab:dns-clusters}). While we see roughly the same number of global providers across hosting and DNS, the
use of managed DNS providers (e.g., Neustar UltraDNS, NSONE) shifts numbers towards
large global providers. Both Neustar UltraDNS and NSONE are seen in the top ten
providers for more than a hundred countries. We also observe a shift away from small regional providers towards large regional providers. In Czechia (the least centralized country)
the share of websites using large regional providers goes up from 39\% (hosting)
to 47\% (DNS).
Interestingly, this shift towards large global providers does not 
markedly increase centralization (\csregion{DNS} = 0.1379, \csregion{Host} = 0.1429) given the reliance on multiple providers. 

%



\dnsclasstable



The shift towards larger providers has led to an increased
anycast use for nameservice infrastructure compared to hosting
(Figure~\ref{fig:dns_geo_region_continent}). While we see global adoption,  
Europe lags behind in anycast
adoption, likely due to heavy reliance on regional
providers in Europe who do not employ it. 






%% file: country_tld.tex
\section{Top Level Domains (TLDs)}
\label{sec:country_tld}

\cdfinsularity

\csregionboxplot
\insularityboxplot

Overall, there is greater centralization in countries' use of TLDs ($\bar{\mathscr{S}}$ = 0.3262) compared to other layers  (Figure~\ref{fig:cs_boxplots_regions}). Despite a large number of top level domains, TLD centralization is primarily driven by global usage of \dns{.com} and countries' usage of ccTLDs. In the most centralized case, 77\% of the top websites in the U.S (\cs = 0.5853) use the \dns{.com} TLD, while in the the least centralized case, Kyrgyzstan (\cs = 0.1468), \dns{.com} accounts for 29\% of the top websites while \dns{.ru} and \dns{.kg} account for 22\% and 12\% respectively. The top four most centralized countries, all in North America (\csavg = 0.4930) and Caribbean (\csavg = 0.4042), all rise to the top because of their \dns{.com} usage. 

After the handful of top countries that rely heavily on \dns{.com}, we begin to see countries rise to be most centralized based on their usage of a ccTLD\@. While there are a few exceptions (\eg, use of \dns{.ru} by CIS countries), most countries centralize on their own ccTLDs. Eastern Europe (\csavg = 0.3361) stands out with Czechia (\cs = 0.4656), Hungary (\cs = 0.4450), and Poland (\cs = 0.4265) being the 5th--7th most centralized countries due to their reliance on their local ccTLDs. We also see that countries tend to be most insular at the TLD layer (Figure~\ref{fig:insularity_cdf}). This may be because the choice of TLD has relatively few technical or financial implications compared to other network layers, providing operators more flexibility to choose an in-country option.


\insularitytldbarplot

When we look at insularity across layers and regions, we see two classes of behavior (Figure~\ref{fig:insularity_boxplots_regions}). First, countries in Europe, Eastern Asia (Japan and Korea), and North America are consistently the most insular countries across all layers. In these countries, the use of local ccTLD is also coupled with a strong network of regional providers (\eg, Japan, Czechia). As such, if a country overcomes the infrastructure barrier and is insular in its hosting infrastructure, we expect it to strongly indicate insularity in the TLD layer. In fact, insularity in the hosting layer is moderately correlated with insularity in the TLD layer ($\rho$=0.70, p $\ll$ 0.05). In the second case, countries in the Global South show insularity most directly at the TLD layer but have low insularity in other layers since equivalent providers do not exist in-country (or, in some cases, even in-region).

Finally, when we look at countries with a large external dependency on their ccTLD, we see evidence of sociopolitical relationships that have led countries to depend on a few countries across different layers: 

\paragraph{France}
Similar to regional clusters in hosting, we find
the \dns{.fr} ccTLD is commonly used across 14 countries: Burkina Faso, Benin, Congo, Côte d'Ivoire, Cameroon,
Algeria, Guadeloupe, Haiti, Madagascar, Mali, Martinique, Réunion, Senegal,
and Togo, where \dns{.fr} is also more popular than their own local ccTLDs. 

\paragraph{Russia} 
Similar to hosting providers, we see the dependence of some old Soviet states
--- countries that are now part of the Commonwealth of Independent States (CIS)
on \dns{.ru}. Not only do these countries use \dns{.ru} but also their own ccTLD
in addition to \dns{.com} making them the least centralized countries given their
lack of overt use of a single TLD.

\paragraph{Germany}
Several countries where German is a dominant language use the \dns{.de} ccTLD.
These include Germany (44\%), Austria (14\%), Luxembourg (8\%), and
Switzerland (7\%). This observation is consistent with prior work~\cite{noms22_hosting_consolidation}.

%% file: country_ca.tex
\section{Certificate Authorities}
\label{sec:country_ca}




\begin{table}[]
\small
\begin{tabular}{lrll}
\toprule
\textbf{Class} & \multicolumn{1}{c}{\textbf{Count}} & \textbf{Description} & \textbf{Example} \\
\midrule
L-GP & 7 & Large Global & DigiCert \\
M-GP & 2 & Medium Global & Entrust \\
L-RP & 11 & Large Regional & Asseco \\
S-RP & 10 & Small Regional & SSL.com \\
XS-RP & 15 & Extra Small Regional & TrustCor \\
\bottomrule
\end{tabular}
    \caption{Classes of Certificate Authorities (CAs)---\textnormal{CAs clustered based on their use and endemicity similar to clustering used for hosting providers.}}
    \label{tab:ca-clusters}
\end{table}


\cshistogramall

Websites' dependencies on certificate authorities follows dramatically different centralization and regionalization patterns than other network dependencies (Figures~\ref{fig:cs_boxplots_regions}~and~\ref{fig:insularity_boxplots_regions}). Unlike the tens of thousands of hosting and DNS
providers that an operator can choose from, we find only 45~CAs used by the websites in our dataset. As a result, centralization on a small number of CAs is consistent across all countries:
$\bar{\mathscr{S}}$ = 0.2007, var = 0.0007
(Figure~\ref{fig:centralization_score_histograms}~and~\ref{fig:centralization_ca_barplot}). However, even within the small set of CAs, use is heavily skewed to a handful of large
authorities: seven CAs account for 98\% of websites. 

At a high level, we see that a handful of large global CAs (e.g., Let's Encrypt) dominate the ecosystem with a few regional CAs such as Asseco (a Polish CA) being heavily used in select countries. In the most centralized case, Slovakia (\cs = 0.3304), 
Let's Encrypt accounts for 55\% of websites, three CAs account for 97\% of websites, and seven CAs account for 98\% of websites ({Appendix~\ref{subappendix:stacked_rest}, Figure \ref{fig:hosting_ca_stacked}). 
The next most centralized countries, Czechia (\cs = 0.3268) 
and Estonia (\cs = 0.2811), show similar distributions. Potentially exacerbating centralization, hosting providers often partner with CAs to issue
certificates for hosted websites. For example, Cloudflare uses Let's Encrypt,
DigiCert, Google, and Sectigo while Incapsula (Imperva) uses
GlobalSign~\cite{cloudflare_cas,imperva_cas}.

Similar to hosting and DNS, we classify CAs based on usage and endemicity, identifying five classes (Table~\ref{tab:ca-clusters}).
Based on this classification, the most important class of CAs is L-GP, which
includes the seven large global CAs that together dominate the web: {Let's Encrypt, DigiCert,
Sectigo, Google, Amazon, GlobalSign, GoDaddy}.
These seven global providers account for 80--99.7\% of websites (Iran and Russia, respectively, at the extremes). DigiCert and Let's Encrypt are the two most popular CAs
accounting for most of the use. 
In the least centralized cases, Taiwan (\cs = 0.1308) and Japan (\cs = 0.1499),
the seven large global CAs account for 82\% and 85\% of websites respectively;
the remaining sites are secured by regional CAs. 
We note that Let's Encrypt is heavily used in European
countries, especially Eastern European countries that use regional hosting
providers. Interestingly, Russia shows the most use of global providers even
though it is one of the most insular countries in other layers. This is consistent with prior work that has
observed that Russia heavily depends on Let's Encrypt and GlobalSign once DigiCert
largely moved out of Russia following the full-scale invasion of Ukraine in
2022~\cite{imc22_whereru}.

\subsection{Regionalization and Insularity}

Because most countries do not have regional CAs, insularity is near zero in all but a handful of places. This leads to the countries that had the highest insularity in other layers to be most different from the CA layer. For example, while Europe had 7 out of the 10 \textit{least} centralized countries at the hosting layer, it has 8 of the 10 \textit{most} centralized countries in the CA layer. Czechia and Slovakia, which place
among the least centralized at the hosting (\cscc{CZ} = 0.0561, \cscc{SK} =
0.0497) and DNS layers (\cscc{CZ} = 0.0391, \cscc{SK} = 0.0429) are the most
centralized from the perspective of CAs (\cscc{CZ} = 0.3268, \cscc{SK} =
0.3304). More generally, Europe, which tends to be more insular in the hosting (Figure~\ref{fig:insularity_barplot}), DNS (Figure~\ref{fig:insularity_dns_barplot}), and
TLD layers (Figure~\ref{fig:insularity_tld_barplot}),
tends to not only be less insular (Figure~\ref{fig:insularity_ca_barplot}) but also more centralized ($\bar{\mathscr{S}}_{EU}$
= 0.2220) at the CA layer. The most popular regional CA is a Polish CA, Asseco (previously Unizeto
Centrum). Asseco is not only used in Poland (19\%) but also in
Iran (19\%) and Afghanistan (5\%) which may be due to these websites avoiding US based CAs. With the exception of Iran and Afghanistan, the use of regional CAs is concentrated in their home country. 
Poland, Taiwan, and Japan are the most insular countries, with respectively
19\%, 17\%, and 14\% of the websites using local CAs.

Centralization on a small number of CAs is not inherently bad for the overall security of the Internet since it reduces the attack surface of the WebPKI\@. However, it also means that the vast majority of countries are dependent on infrastructure in the U.S. and have few alternatives to choose from. Recently, several countries and regions have tried to change this. When major western CAs left Russia following the invasion of Ukraine~\cite{eff_russian_root_ca}, Russia attempted to stand up a state-sponsored Root CA in 2022, but the root certificate was never accepted by major web browsers. The European Union also recently attempted to pass regulations on how trust decisions are made by web browsers, which was met with significant backlash from both web browsers and the broader security community~\cite{eu_qwacs,record_eidas}.



\insularitycabarplot



%
%
%

%% file: discussion.tex
\section{Discussion}

We hope that quantifying both centralization and regionalization enables the
networking community to better understand and discuss how the Internet is
structured and changing over time. In this work, we introduced new quantitative centralization, usage, and endemicity metrics, which we used to look at centralization across countries, shedding light on the nuances of web centralization and how regional undercurrents affect the experiences of users on the web. Here, we highlight implications and open questions from our work.



\vspace{2pt}
\paragraph{The Role of Regional Providers} Cloudflare (and to
a lesser degree other large global providers) significantly impacts 
the centralization of the web. However, when we focus our discussion on the size of providers---especially global providers---it is easy to lose the context of how centralization
impacts people's lived experiences on the web. A country can be significantly centralized on a single local provider, but that provider could appear small on the global stage. For example, SuperHosting.BG in Bulgaria and UAB in Lithuania have not received substantial attention in
prior discussion of centralization despite driving centralization in their respective countries. On the other hand, large global providers like Google, Akamai and Fastly are used by many countries, but only modestly affect centralization for people in each country.  Even in the common case where a country's use of a range of regional providers contributes to a more
diffuse ecosystem, the vast majority of the websites
still converge on the top~10 or so providers.  For instance, in Iran, the least centralized country, 60\% of sites are still served by only ten providers. 
As such, while our centralization
score provides a meaningful scale for comparing countries, the underlying nature
of providers and distribution is still important.

\vspace{2pt}
\paragraph{Interplay Between Layers}
Looking at provider usage across layers identifies dependencies that may be driving centralization. For instance, the use
of a hosting provider is at times tightly coupled with a specific DNS
provider. For example, Cloudflare's service is predicated on using their DNS
service in most cases. Moreover, Cloudflare also allows websites using
them to obtain a TLS certificate from four different CAs (DigiCert, Let's Encrypt, Google, and Sectigo)~\cite{cloudflare_cas},
which also happen to be four of the seven large global CAs that
dominate the web (Section~\ref{sec:country_ca}).
Similarly, other hosting providers such as Incapsula
(Imperva) also secure websites with one CA: GlobalSign~\cite{imperva_cas}. 
Thus, it is likely that much of the centralization we see on the web is a
result of choices made by providers as opposed to an explicit choice made
by the website operator themselves.
%
As such, future work should analyze these covariates to
establish causation behind the patterns we observe.



\vspace{2pt}
\paragraph{Root Causes of Centralization} By examining centralization both
geographically and across layers, our work uncovers patterns that
hint at some of the forces that may drive centralization. 
For instance, based on
observed regional dependence, we might hypothesize that historical and
linguistic ties contribute to countries coalescing around a particular set of
providers, and it is reasonable to conjecture that the reason Iran has
centralized on Let's Encrypt is different from the reasons why Thailand has
centralized on Cloudflare. Perhaps more concretely, we may argue sanctions drove Russia's particular
dependence on Let's Encrypt and GlobalSign~\cite{imc22_whereru}. However, this work does not provide a true causal
analysis: there is much still to be done to understand \textit{why} we see the
patterns we do. This requires handling challenging confounds: for example, do
Slovakian websites choose to use Czech providers, or do Slovakian people tend to
frequent Czech websites? Do the choices by a provider, such as a
hosting provider's default CA configuration or DNS offering, cause a synergy in
centralization patterns across infrastructure layers? Exploring these complexities is key to understanding and affecting the future of centralization.


\vspace{2pt}
\paragraph{The Role of Regionalization}
Centralization is important because it represents a form of \textit{dependence} that---for better or for worse---affects users and the websites they are trying to reach. However, centralization captures only certain forms of dependence and notably may overlook regional dependencies. For instance, CIS countries tend to exhibit less centralization on hosting providers but still depend heavily on Russian providers. Similarly, many countries in Africa rely on providers in France due to various sociopolitical reasons (\eg, colonial ties). 
While these regional providers may not be the largest providers globally, from the perspective of these countries, dependence on a set of providers in a single external country carries similar risks to that of centralization on large global providers. 
It is in this context of dependence that we stress the importance of considering
both regional dependence and provider dependence when studying the web.
Together, these dimensions allow us to paint a more holistic picture of dependence on the web, enabling us to make informed decisions on mitigating risks associated with geopolitical influences on the Internet.

%


\section{conclusion}
In conclusion, our work demonstrates the intricate interplay between centralization, regionalization, and insularity across various layers of internet infrastructure. By analyzing the influence of both global and regional providers, we highlight the crucial role that geopolitical and sociopolitical factors play in shaping the web landscape. Moving forward, we hope that the quantitative metrics we use to measure centralization and regionalization on the web enables future studies to more deeply understand the Internet and how it is evolving.

%% file: emd_code.tex
\section{Ethics}

Our work does not involve human subjects and therefore, according to our institution’s IRB policies, does not require IRB approval. On the server we use for performing active measurements, we follow the active scanning guidelines outlined by Durumeric \etal~\cite{durumeric2013zmap} when actively identifying server purpose and ownership. We did not receive any opt-out requests. 

The centralization of the Internet has been a topic of significant debate. In this work, we seek to avoid making a value judgement on Internet centralization, but rather, to provide the research community with a suite of tools by which to measure centralization and make observations on how Internet centralization is operationalized today. We also stress that we are a team of researchers located in a western country discussing the Internet architecture of many countries and world-regions we are not a part of. A team of researchers in other world regions and shaped by different cultural contexts may experience Internet centralization in different and even personal ways. As such, our identities both shape and potentially limit our findings.

\section{EMD Reference}
\label{sec:emd}
\label{sec:emd:math}

Here, we provide additional details about how EMD is formally defined and how we arrive at a simple expression for our instantiation of it. We begin with a general definition of EMD in its discretized formalization.

Let $A = (a_1, a_2, \ldots, a_n)$ be the observed distribution of data, and let $R = (r_1, r_2, \ldots, r_m)$ be the reference distribution. For simplicity, assume $\sum_{i=1}^{n} a_i = \sum_{j=1}^{m} r_j$. We define a ground distance function $d_{ij}$ for all $1 \le i \le n$, $1 \le j \le m$. Transforming $A$ into $R$ requires assigning nonnegative \textit{flow} values $f_{ij}$ for all $i, j$, which define the amount of earth moved from bucket $i$ to bucket $j$. Naturally, the total flow out of a pile equals amount originally in the pile ($\sum_{j=1}^{m} f_{ij} = a_i \ \forall i$), and the total flow into a pile equals the ending size of the pile ($\sum_{i=1}^{n} f_{ij} = r_j \ \forall j$). There may be many solutions satisfying these constraints, but we aim to find one that minimizes the total \textit{work} involved in transporting the earth, which is defined as the product of flow and distance:
$$ w = \min_{[f_{ij}]} \sum_{i=1}^{n} \sum_{j=1}^{m} f_{ij} d_{ij} $$
Taking the $f_{ij}$ that solve this minimization problem, the EMD is defined as this minimum work, optionally normalized to $[0, 1]$ by means appropriate to the chosen distance metric and total flow. If $0 \le d_{ij} \le 1 \ \forall i,j$ above, then the expression becomes: 
$$ EMD(A, R) = \frac{\sum_{i=1}^{n} \sum_{j=1}^{m} f_{ij} d_{ij}}{\sum_{i=1}^{n} \sum_{j=1}^{m} f_{ij}} = \frac{\sum_{i=1}^{n} \sum_{j=1}^{m} f_{ij} d_{ij}}{\sum_{i=1}^{n} a_i}$$

\noindent For our instantiation of EMD in this paper, we can use our uniform reference distribution and simple vertical-difference distance function to enable further simplification. Specifically: 
\begin{itemize}
    \item Let $A = (a_1, a_2, \ldots, a_n)$ be a nonincreasing sequence representing the counts of websites using each provider, with $\sum_i a_i = C$ the total number of sites.
    \item Let $R$ be a uniform distribution across $C$ buckets, each of size 1, representing a fully decentralized reference distribution.
    \item Let $d_{ij} = \frac{1}{C} (a_i - r_j) = \frac{a_i - 1}{C} \ \forall i,j$. This represents the vertical height difference between $a_i$ and $r_j$, normalized by the total number of sites.
\end{itemize}

\noindent Next, we determine the optimum flow. For each $i$, we need to move each of $a_i$ units of size 1 (each website) into a separate bucket in $R$. (It does not matter which, as $d_{ij}$ is only dependent on $i$, not $j$.) So for each $i$ we have $f_{ij} = 1$ for $a_i$ values of $j$ and $f_{ij} = 0$ for all other $j$. This means the total work is
$$w = \sum_{i=0}^n a_i (1 \cdot \frac{a_i - 1}{C}) = \sum_{i=0}^n \frac{a_i^2 - a_i}{C} = \sum_{i=0}^n \frac{a_i^2}{C} - \frac{\sum_{i=0}^n a_i}{C} = \sum_{i=0}^n \frac{a_i^2}{C} - 1$$
To normalize to $[0,1]$, we divide by the total flow, which also equals $C$. This yields our final expression for our centralization score \cs:
$$\textnormal{\cs} = \frac{1}{C} \left( \sum_{i=0}^n \frac{a_i^2}{C} - 1 \right) = \sum_{i=0}^n\left(\frac{a_i}{C}\right)^2 - \frac{1}{C}$$

%% file: country_references.tex
\section{Country Reference}
\label{app:country_references}

Table~\ref{tab:country_ref} lists all 150 countries included in our dataset.

\begin{table}
\scriptsize
\renewcommand{\arraystretch}{0.95}
\begin{tabular}{llll|llll}
    \toprule
\multicolumn{1}{c}{\textbf{CC}} & \multicolumn{1}{c}{\textbf{Country}} & \multicolumn{1}{c}{\textbf{Region}} & \multicolumn{1}{c}{\textbf{Continent}} & \multicolumn{1}{c}{\textbf{CC}} & \multicolumn{1}{c}{\textbf{Country}} & \multicolumn{1}{c}{\textbf{Region}} & \multicolumn{1}{c}{\textbf{Continent}} \\
\midrule
AE & United Arab Emirates & Western Asia & AS & LT & Lithuania & Northern Europe & EU \\ 
AF & Afghanistan & Southern Asia & AS & LU & Luxembourg & Western Europe & EU \\ 
AL & Albania & Southern Europe & EU & LV & Latvia & Northern Europe & EU \\ 
AM & Armenia & Western Asia & AS & LY & Libya & Northern Africa & AF \\ 
AO & Angola & Middle Africa & AF & MA & Morocco & Northern Africa & AF \\ 
AR & Argentina & South America & SA & MD & Moldova & Eastern Europe & EU \\ 
AT & Austria & Western Europe & EU & ME & Montenegro & Southern Europe & EU \\ 
AU & Australia & Oceania & OC & MG & Madagascar & Eastern Africa & AF \\ 
AZ & Azerbaijan & Western Asia & AS & MK & North Macedonia & Southern Europe & EU \\ 
BA & Bosnia and Herzegovina & Southern Europe & EU & ML & Mali & Western Africa & AF \\ 
BD & Bangladesh & Southern Asia & AS & MM & Myanmar & South-eastern Asia & AS \\ 
BE & Belgium & Western Europe & EU & MN & Mongolia & Eastern Asia & AS \\ 
BF & Burkina Faso & Western Africa & AF & MO & Macao & Eastern Asia & AS \\ 
BG & Bulgaria & Eastern Europe & EU & MQ & Martinique & Caribbean & NA \\ 
BH & Bahrain & Western Asia & AS & MT & Malta & Southern Europe & EU \\ 
BJ & Benin & Western Africa & AF & MU & Mauritius & Eastern Africa & AF \\ 
BN & Brunei Darussalam & South-eastern Asia & AS & MV & Maldives & Southern Asia & AS \\ 
BO & Bolivia & South America & SA & MW & Malawi & Eastern Africa & AF \\ 
BR & Brazil & South America & SA & MX & Mexico & Central America & NA \\ 
BW & Botswana & Southern Africa & AF & MY & Malaysia & South-eastern Asia & AS \\ 
BY & Belarus & Eastern Europe & EU & MZ & Mozambique & Eastern Africa & AF \\ 
CA & Canada & Northern America & NA & NA & Namibia & Southern Africa & AF \\ 
CD & Congo & Middle Africa & AF & NG & Nigeria & Western Africa & AF \\ 
CH & Switzerland & Western Europe & EU & NI & Nicaragua & Central America & NA \\ 
CI & Côte d'Ivoire & Western Africa & AF & NL & Netherlands & Western Europe & EU \\ 
CL & Chile & South America & SA & NO & Norway & Northern Europe & EU \\ 
CM & Cameroon & Middle Africa & AF & NP & Nepal & Southern Asia & AS \\ 
CO & Colombia & South America & SA & NZ & New Zealand & Oceania & OC \\ 
CR & Costa Rica & Central America & NA & OM & Oman & Western Asia & AS \\ 
CU & Cuba & Caribbean & NA & PA & Panama & Central America & NA \\ 
CY & Cyprus & Western Asia & AS & PE & Peru & South America & SA \\ 
CZ & Czechia & Eastern Europe & EU & PG & Papua New Guinea & Oceania & OC \\ 
DE & Germany & Western Europe & EU & PH & Philippines & South-eastern Asia & AS \\ 
DK & Denmark & Northern Europe & EU & PK & Pakistan & Southern Asia & AS \\ 
DO & Dominican Republic & Caribbean & NA & PL & Poland & Eastern Europe & EU \\ 
DZ & Algeria & Northern Africa & AF & PR & Puerto Rico & Caribbean & NA \\ 
EC & Ecuador & South America & SA & PS & Palestine & Western Asia & AS \\ 
EE & Estonia & Northern Europe & EU & PT & Portugal & Southern Europe & EU \\ 
EG & Egypt & Northern Africa & AF & PY & Paraguay & South America & SA \\ 
ES & Spain & Southern Europe & EU & QA & Qatar & Western Asia & AS \\ 
ET & Ethiopia & Eastern Africa & AF & RE & Réunion & Eastern Africa & AF \\ 
FI & Finland & Northern Europe & EU & RO & Romania & Eastern Europe & EU \\ 
FR & France & Western Europe & EU & RS & Serbia & Southern Europe & EU \\ 
GA & Gabon & Middle Africa & AF & RU & Russia & Eastern Europe & EU \\ 
GB & United Kingdom & Northern Europe & EU & RW & Rwanda & Eastern Africa & AF \\ 
GE & Georgia & Western Asia & AS & SA & Saudi Arabia & Western Asia & AS \\ 
GH & Ghana & Western Africa & AF & SD & Sudan & Northern Africa & AF \\ 
GP & Guadeloupe & Caribbean & NA & SE & Sweden & Northern Europe & EU \\ 
GR & Greece & Southern Europe & EU & SG & Singapore & South-eastern Asia & AS \\ 
GT & Guatemala & Central America & NA & SI & Slovenia & Southern Europe & EU \\ 
HK & Hong Kong & Eastern Asia & AS & SK & Slovakia & Eastern Europe & EU \\ 
HN & Honduras & Central America & NA & SN & Senegal & Western Africa & AF \\ 
HR & Croatia & Southern Europe & EU & SO & Somalia & Eastern Africa & AF \\ 
HT & Haiti & Caribbean & NA & SV & El Salvador & Central America & NA \\ 
HU & Hungary & Eastern Europe & EU & SY & Syria & Western Asia & AS \\ 
ID & Indonesia & South-eastern Asia & AS & TG & Togo & Western Africa & AF \\ 
IE & Ireland & Northern Europe & EU & TH & Thailand & South-eastern Asia & AS \\ 
IL & Israel & Western Asia & AS & TJ & Tajikistan & Central Asia & AS \\ 
IN & India & Southern Asia & AS & TM & Turkmenistan & Central Asia & AS \\ 
IQ & Iraq & Western Asia & AS & TN & Tunisia & Northern Africa & AF \\ 
IR & Iran & Southern Asia & AS & TR & Turkey & Western Asia & AS \\ 
IS & Iceland & Northern Europe & EU & TT & Trinidad and Tobago & Caribbean & NA \\ 
IT & Italy & Southern Europe & EU & TW & Taiwan & Eastern Asia & AS \\ 
JM & Jamaica & Caribbean & NA & TZ & Tanzania & Eastern Africa & AF \\ 
JO & Jordan & Western Asia & AS & UA & Ukraine & Eastern Europe & EU \\ 
JP & Japan & Eastern Asia & AS & UG & Uganda & Eastern Africa & AF \\ 
KE & Kenya & Eastern Africa & AF & US & United States & Northern America & NA \\ 
KG & Kyrgyzstan & Central Asia & AS & UY & Uruguay & South America & SA \\ 
KH & Cambodia & South-eastern Asia & AS & UZ & Uzbekistan & Central Asia & AS \\ 
KR & Korea & Eastern Asia & AS & VE & Venezuela & South America & SA \\ 
KW & Kuwait & Western Asia & AS & VN & Viet Nam & South-eastern Asia & AS \\ 
KZ & Kazakhstan & Central Asia & AS & YE & Yemen & Western Asia & AS \\ 
LA & Laos & South-eastern Asia & AS & ZA & South Africa & Southern Africa & AF \\ 
LB & Lebanon & Western Asia & AS & ZM & Zambia & Eastern Africa & AF \\ 
LK & Sri Lanka & Southern Asia & AS & ZW & Zimbabwe & Eastern Africa & AF \\ 

\bottomrule
\end{tabular}
\caption{Reference for all the country codes and their regions.}
\label{tab:country_ref}
\end{table}

%% file: emd_scores.tex
\section{Centralization Scores for 150 Countries}
\label{sec:cs_scores}

Tables~\ref{tab:country_provider_all_scores},~\ref{tab:country_dns_all_scores},~\ref{tab:country_ca_all_scores}, and~\ref{tab:country_tld_all_scores} show each country's centralization score for the hosting, DNS, CA, and TLD layers, respectively.

\begin{table}[h!]
\small
\begin{tabular}{llll|llll|llll|llll}
\toprule
\multicolumn{1}{c}{\textbf{Rank}} & \multicolumn{2}{c}{\textbf{Country}} & \multicolumn{1}{c}{\textbf{CS}} & \multicolumn{1}{c}{\textbf{Rank}} & \multicolumn{2}{c}{\textbf{Country}} & \multicolumn{1}{c}{\textbf{CS}} & \multicolumn{1}{c}{\textbf{Rank}} & \multicolumn{2}{c}{\textbf{Country}} & \multicolumn{1}{c}{\textbf{CS}} & \multicolumn{1}{c}{\textbf{Rank}} & \multicolumn{2}{c}{\textbf{Country}} & \multicolumn{1}{c}{\textbf{CS}} \\
\midrule
1 & TH & AS & 0.3548 & 41 & JM & NA & 0.1702 & 81 & CM & AF & 0.131 & 121 & DE & EU & 0.0947 \\
2 & ID & AS & 0.3258 & 42 & VN & AS & 0.1694 & 82 & CA & NA & 0.1308 & 122 & NO & EU & 0.0937 \\
3 & MM & AS & 0.2641 & 43 & ZM & AF & 0.1653 & 83 & CR & NA & 0.1287 & 123 & HR & EU & 0.0931 \\
4 & LA & AS & 0.2526 & 44 & AO & AF & 0.1623 & 84 & LT & EU & 0.1286 & 124 & AR & SA & 0.0928 \\
5 & IQ & AS & 0.249 & 45 & GH & AF & 0.1608 & 85 & RW & AF & 0.1275 & 125 & ES & EU & 0.0918 \\
6 & LY & AF & 0.2462 & 46 & MW & AF & 0.1603 & 86 & SN & AF & 0.1273 & 126 & TW & AS & 0.0914 \\
7 & SY & AS & 0.2379 & 47 & IN & AS & 0.16 & 87 & TG & AF & 0.1266 & 127 & RS & EU & 0.0905 \\
8 & PK & AS & 0.23 & 48 & ZA & AF & 0.1549 & 88 & CI & AF & 0.1247 & 128 & AF & AS & 0.0904 \\
9 & KH & AS & 0.2299 & 49 & HN & NA & 0.1545 & 89 & BJ & AF & 0.1244 & 129 & PL & EU & 0.0887 \\
10 & OM & AS & 0.2287 & 50 & NI & NA & 0.1537 & 90 & GA & AF & 0.1232 & 130 & BE & EU & 0.088 \\
11 & SA & AS & 0.2282 & 51 & NZ & OC & 0.1524 & 91 & UA & EU & 0.1228 & 131 & MD & EU & 0.0876 \\
12 & PS & AS & 0.2254 & 52 & MZ & AF & 0.1519 & 92 & CD & AF & 0.1219 & 132 & LV & EU & 0.0873 \\
13 & KW & AS & 0.2228 & 53 & DO & NA & 0.1511 & 93 & PE & SA & 0.1218 & 133 & RO & EU & 0.0869 \\
14 & YE & AS & 0.2219 & 54 & NA & AF & 0.1508 & 94 & CL & SA & 0.1213 & 134 & KG & AS & 0.0868 \\
15 & LB & AS & 0.2219 & 55 & AU & OC & 0.1504 & 95 & MX & NA & 0.1203 & 135 & IT & EU & 0.0859 \\
16 & JO & AS & 0.2198 & 56 & PA & NA & 0.1495 & 96 & ML & AF & 0.1193 & 136 & TJ & AS & 0.0844 \\
17 & SD & AF & 0.2188 & 57 & NG & AF & 0.1493 & 97 & MK & EU & 0.1192 & 137 & CH & EU & 0.0842 \\
18 & NP & AS & 0.2167 & 58 & VE & SA & 0.1488 & 98 & EC & SA & 0.1192 & 138 & MO & AS & 0.0839 \\
19 & QA & AS & 0.2161 & 59 & PR & NA & 0.1478 & 99 & BG & EU & 0.1188 & 139 & KR & AS & 0.0825 \\
20 & EG & AF & 0.2155 & 60 & GB & EU & 0.1463 & 100 & HK & AS & 0.118 & 140 & AT & EU & 0.0816 \\
21 & BH & AS & 0.2151 & 61 & MT & EU & 0.1462 & 101 & RE & AF & 0.114 & 141 & FI & EU & 0.0815 \\
22 & MY & AS & 0.2143 & 62 & CU & NA & 0.1459 & 102 & BA & EU & 0.1121 & 142 & KZ & AS & 0.079 \\
23 & DZ & AF & 0.2126 & 63 & BR & SA & 0.1446 & 103 & AM & AS & 0.1103 & 143 & BY & EU & 0.0766 \\
24 & SG & AS & 0.2003 & 64 & ZW & AF & 0.1443 & 104 & GE & AS & 0.1086 & 144 & SI & EU & 0.0645 \\
25 & SO & AF & 0.1991 & 65 & KE & AF & 0.1431 & 105 & LU & EU & 0.108 & 145 & HU & EU & 0.0604 \\
26 & BN & AS & 0.1983 & 66 & CY & AS & 0.1418 & 106 & FR & EU & 0.1069 & 146 & CZ & EU & 0.0561 \\
27 & BD & AS & 0.1971 & 67 & UG & AF & 0.1406 & 107 & UY & SA & 0.1066 & 147 & RU & EU & 0.0554 \\
28 & AE & AS & 0.1937 & 68 & IE & EU & 0.1398 & 108 & PT & EU & 0.1065 & 148 & SK & EU & 0.0497 \\
29 & PH & AS & 0.1934 & 69 & TZ & AF & 0.1395 & 109 & NL & EU & 0.1062 & 149 & TM & AS & 0.0461 \\
30 & MA & AF & 0.1852 & 70 & TR & AS & 0.1394 & 110 & CO & SA & 0.1044 & 150 & IR & AS & 0.0411 \\
31 & TN & AF & 0.1848 & 71 & SV & NA & 0.1374 & 111 & JP & AS & 0.1036 &  &  &  &  \\
32 & MV & AS & 0.1823 & 72 & MN & AS & 0.136 & 112 & IS & EU & 0.1025 &  &  &  &  \\
33 & AL & EU & 0.1806 & 73 & HT & NA & 0.1359 & 113 & ME & EU & 0.102 &  &  &  &  \\
34 & ET & AF & 0.1764 & 74 & PY & SA & 0.1359 & 114 & SE & EU & 0.1018 &  &  &  &  \\
35 & TT & NA & 0.1755 & 75 & US & NA & 0.1358 & 115 & BF & AF & 0.1018 &  &  &  &  \\
36 & PG & OC & 0.1755 & 76 & GT & NA & 0.134 & 116 & GP & NA & 0.1011 &  &  &  &  \\
37 & LK & AS & 0.1749 & 77 & BO & SA & 0.1335 & 117 & DK & EU & 0.101 &  &  &  &  \\
38 & AZ & AS & 0.1743 & 78 & IL & AS & 0.132 & 118 & MQ & NA & 0.1007 &  &  &  &  \\
39 & MU & AF & 0.1737 & 79 & GR & EU & 0.1319 & 119 & UZ & AS & 0.0978 &  &  &  &  \\
40 & BW & AF & 0.1727 & 80 & MG & AF & 0.1318 & 120 & EE & EU & 0.097 &  &  &  & \\
\bottomrule
\end{tabular}
\caption{Country x Provider Centralization Scores}
\label{tab:country_provider_all_scores}
\end{table}

\begin{table}[h!]
\small
\begin{tabular}{llll|llll|llll|llll}
\toprule
\multicolumn{1}{c}{\textbf{Rank}} & \multicolumn{2}{c}{\textbf{Country}} & \multicolumn{1}{c}{\textbf{CS}} & \multicolumn{1}{c}{\textbf{Rank}} & \multicolumn{2}{c}{\textbf{Country}} & \multicolumn{1}{c}{\textbf{CS}} & \multicolumn{1}{c}{\textbf{Rank}} & \multicolumn{2}{c}{\textbf{Country}} & \multicolumn{1}{c}{\textbf{CS}} & \multicolumn{1}{c}{\textbf{Rank}} & \multicolumn{2}{c}{\textbf{Country}} & \multicolumn{1}{c}{\textbf{CS}} \\
\midrule
1 & ID & AS & 0.3757 & 41 & JM & NA & 0.1712 & 81 & EC & SA & 0.1227 & 121 & LU & EU & 0.0808 \\
2 & TH & AS & 0.3374 & 42 & MY & AS & 0.1700 & 82 & US & NA & 0.1221 & 122 & FR & EU & 0.0805 \\
3 & IQ & AS & 0.2730 & 43 & ZM & AF & 0.1651 & 83 & CO & SA & 0.1214 & 123 & KR & AS & 0.0804 \\
4 & SY & AS & 0.2653 & 44 & MU & AF & 0.1643 & 84 & MK & EU & 0.1212 & 124 & GP & NA & 0.0797 \\
5 & LY & AF & 0.2548 & 45 & DO & NA & 0.1628 & 85 & SN & AF & 0.1189 & 125 & MQ & NA & 0.0793 \\
6 & MM & AS & 0.2469 & 46 & NI & NA & 0.1624 & 86 & UY & SA & 0.1179 & 126 & NL & EU & 0.0793 \\
7 & SD & AF & 0.2439 & 47 & NG & AF & 0.1611 & 87 & TG & AF & 0.1173 & 127 & DK & EU & 0.0792 \\
8 & NP & AS & 0.2430 & 48 & VE & SA & 0.1610 & 88 & AM & AS & 0.1168 & 128 & TW & AS & 0.0775 \\
9 & YE & AS & 0.2346 & 49 & GH & AF & 0.1607 & 89 & BJ & AF & 0.1164 & 129 & HR & EU & 0.0774 \\
10 & PS & AS & 0.2340 & 50 & MW & AF & 0.1601 & 90 & MG & AF & 0.1157 & 130 & HK & AS & 0.0760 \\
11 & OM & AS & 0.2340 & 51 & HN & NA & 0.1600 & 91 & BG & EU & 0.1155 & 131 & PL & EU & 0.0760 \\
12 & BD & AS & 0.2317 & 52 & BW & AF & 0.1594 & 92 & GE & AS & 0.1142 & 132 & RO & EU & 0.0704 \\
13 & EG & AF & 0.2291 & 53 & AO & AF & 0.1553 & 93 & GA & AF & 0.1135 & 133 & RS & EU & 0.0703 \\
14 & JO & AS & 0.2281 & 54 & CU & NA & 0.1549 & 94 & MX & NA & 0.1124 & 134 & IT & EU & 0.0676 \\
15 & LA & AS & 0.2281 & 55 & GT & NA & 0.1531 & 95 & CD & AF & 0.1123 & 135 & IS & EU & 0.0660 \\
16 & SA & AS & 0.2241 & 56 & PY & SA & 0.1517 & 96 & CI & AF & 0.1119 & 136 & DE & EU & 0.0656 \\
17 & KW & AS & 0.2217 & 57 & MZ & AF & 0.1499 & 97 & ZA & AF & 0.1113 & 137 & NO & EU & 0.0644 \\
18 & DZ & AF & 0.2159 & 58 & BR & SA & 0.1472 & 98 & CA & NA & 0.1099 & 138 & MO & AS & 0.0625 \\
19 & SO & AF & 0.2157 & 59 & SG & AS & 0.1466 & 99 & JP & AS & 0.1097 & 139 & BE & EU & 0.0624 \\
20 & QA & AS & 0.2140 & 60 & KE & AF & 0.1461 & 100 & CL & SA & 0.1072 & 140 & IR & AS & 0.0620 \\
21 & LB & AS & 0.2139 & 61 & PA & NA & 0.1457 & 101 & GB & EU & 0.1072 & 141 & CH & EU & 0.0611 \\
22 & BH & AS & 0.2136 & 62 & SV & NA & 0.1456 & 102 & ML & AF & 0.1052 & 142 & SE & EU & 0.0556 \\
23 & KH & AS & 0.2136 & 63 & UG & AF & 0.1451 & 103 & AF & AS & 0.1047 & 143 & RU & EU & 0.0556 \\
24 & PK & AS & 0.2115 & 64 & TR & AS & 0.1444 & 104 & EE & EU & 0.1001 & 144 & AT & EU & 0.0543 \\
25 & MN & AS & 0.2115 & 65 & CY & AS & 0.1393 & 105 & ME & EU & 0.0966 & 145 & SI & EU & 0.0485 \\
26 & LK & AS & 0.1956 & 66 & BO & SA & 0.1359 & 106 & AR & SA & 0.0953 & 146 & TM & AS & 0.0460 \\
27 & LT & EU & 0.1919 & 67 & HT & NA & 0.1354 & 107 & UA & EU & 0.0953 & 147 & FI & EU & 0.0459 \\
28 & PH & AS & 0.1900 & 68 & TZ & AF & 0.1352 & 108 & UZ & AS & 0.0924 & 148 & SK & EU & 0.0429 \\
29 & BN & AS & 0.1892 & 69 & NA & AF & 0.1342 & 109 & MD & EU & 0.0907 & 149 & HU & EU & 0.0404 \\
30 & AL & EU & 0.1855 & 70 & PE & SA & 0.1332 & 110 & IE & EU & 0.0897 & 150 & CZ & EU & 0.0391 \\
31 & AE & AS & 0.1827 & 71 & NZ & OC & 0.1327 & 111 & BA & EU & 0.0894 &  &  &  &  \\
32 & MV & AS & 0.1817 & 72 & MT & EU & 0.1321 & 112 & RE & AF & 0.0894 &  &  &  &  \\
33 & TT & NA & 0.1805 & 73 & ZW & AF & 0.1305 & 113 & BF & AF & 0.0893 &  &  &  &  \\
34 & TN & AF & 0.1803 & 74 & RW & AF & 0.1300 & 114 & TJ & AS & 0.0868 &  &  &  &  \\
35 & ET & AF & 0.1796 & 75 & PR & NA & 0.1287 & 115 & KG & AS & 0.0862 &  &  &  &  \\
36 & AZ & AS & 0.1772 & 76 & CR & NA & 0.1286 & 116 & BY & EU & 0.0841 &  &  &  &  \\
37 & VN & AS & 0.1769 & 77 & IL & AS & 0.1284 & 117 & ES & EU & 0.0836 &  &  &  &  \\
38 & IN & AS & 0.1755 & 78 & GR & EU & 0.1266 & 118 & PT & EU & 0.0819 &  &  &  &  \\
39 & MA & AF & 0.1750 & 79 & CM & AF & 0.1246 & 119 & KZ & AS & 0.0818 &  &  &  &  \\
40 & PG & OC & 0.1732 & 80 & AU & OC & 0.1235 & 120 & LV & EU & 0.0813 &  &  &  & \\
\bottomrule
\end{tabular}
\caption{Country x DNS Infrastructure Centralization Scores}
\label{tab:country_dns_all_scores}
\end{table}

\begin{table}[h!]
\begin{tabular}{llll|llll|llll|llll}
    \toprule
\multicolumn{1}{c}{\textbf{Rank}} & \multicolumn{2}{c}{\textbf{Country}} & \multicolumn{1}{c}{\textbf{CS}} & \multicolumn{1}{c}{\textbf{Rank}} & \multicolumn{2}{c}{\textbf{Country}} & \multicolumn{1}{c}{\textbf{CS}} & \multicolumn{1}{c}{\textbf{Rank}} & \multicolumn{2}{c}{\textbf{Country}} & \multicolumn{1}{c}{\textbf{CS}} & \multicolumn{1}{c}{\textbf{Rank}} & \multicolumn{2}{c}{\textbf{Country}} & \multicolumn{1}{c}{\textbf{CS}} \\
\midrule
1 & SK & EU & 0.3304 & 41 & IQ & AS & 0.2054 & 81 & HT & NA & 0.1945 & 121 & AZ & AS & 0.1863 \\
2 & CZ & EU & 0.3268 & 42 & MG & AF & 0.2051 & 82 & TN & AF & 0.1943 & 122 & EG & AF & 0.1859 \\
3 & EE & EU & 0.2811 & 43 & IE & EU & 0.2043 & 83 & MW & AF & 0.1943 & 123 & NI & NA & 0.1853 \\
4 & IR & AS & 0.2807 & 44 & PR & NA & 0.2041 & 84 & BF & AF & 0.1937 & 124 & HK & AS & 0.1852 \\
5 & SI & EU & 0.2623 & 45 & MK & EU & 0.2039 & 85 & PS & AS & 0.1937 & 125 & AR & SA & 0.1850 \\
6 & HU & EU & 0.2555 & 46 & FI & EU & 0.2038 & 86 & AM & AS & 0.1936 & 126 & GT & NA & 0.1848 \\
7 & RU & EU & 0.2474 & 47 & ME & EU & 0.2035 & 87 & CY & AS & 0.1932 & 127 & HN & NA & 0.1845 \\
8 & TM & AS & 0.2462 & 48 & ID & AS & 0.2035 & 88 & KW & AS & 0.1930 & 128 & PA & NA & 0.1833 \\
9 & BY & EU & 0.2418 & 49 & BN & AS & 0.2032 & 89 & DZ & AF & 0.1928 & 129 & BO & SA & 0.1828 \\
10 & LT & EU & 0.2404 & 50 & MV & AS & 0.2030 & 90 & UG & AF & 0.1926 & 130 & ES & EU & 0.1816 \\
11 & UA & EU & 0.2354 & 51 & AF & AS & 0.2030 & 91 & IT & EU & 0.1924 & 131 & UY & SA & 0.1810 \\
12 & LV & EU & 0.2332 & 52 & TT & NA & 0.2022 & 92 & CI & AF & 0.1923 & 132 & BD & AS & 0.1804 \\
13 & TJ & AS & 0.2331 & 53 & LU & EU & 0.2020 & 93 & GH & AF & 0.1922 & 133 & CR & NA & 0.1798 \\
14 & MD & EU & 0.2329 & 54 & AL & EU & 0.2012 & 94 & PT & EU & 0.1920 & 134 & SV & NA & 0.1795 \\
15 & GR & EU & 0.2323 & 55 & GB & EU & 0.2012 & 95 & QA & AS & 0.1920 & 135 & VE & SA & 0.1786 \\
16 & KZ & AS & 0.2289 & 56 & DE & EU & 0.2005 & 96 & AO & AF & 0.1920 & 136 & BR & SA & 0.1779 \\
17 & RS & EU & 0.2259 & 57 & LY & AF & 0.2004 & 97 & SN & AF & 0.1918 & 137 & NG & AF & 0.1779 \\
18 & TH & AS & 0.2243 & 58 & GA & AF & 0.1996 & 98 & BH & AS & 0.1917 & 138 & MX & NA & 0.1750 \\
19 & KG & AS & 0.2235 & 59 & MO & AS & 0.1995 & 99 & NA & AF & 0.1917 & 139 & EC & SA & 0.1745 \\
20 & HR & EU & 0.2222 & 60 & TZ & AF & 0.1992 & 100 & ML & AF & 0.1913 & 140 & MN & AS & 0.1738 \\
21 & BG & EU & 0.2200 & 61 & JM & NA & 0.1988 & 101 & GE & AS & 0.1910 & 141 & PH & AS & 0.1738 \\
22 & RO & EU & 0.2198 & 62 & JO & AS & 0.1984 & 102 & BE & EU & 0.1910 & 142 & CL & SA & 0.1683 \\
23 & AT & EU & 0.2183 & 63 & BW & AF & 0.1978 & 103 & PK & AS & 0.1908 & 143 & IN & AS & 0.1683 \\
24 & AU & OC & 0.2179 & 64 & BJ & AF & 0.1976 & 104 & ZM & AF & 0.1907 & 144 & PE & SA & 0.1657 \\
25 & DK & EU & 0.2165 & 65 & SY & AS & 0.1975 & 105 & ET & AF & 0.1903 & 145 & TR & AS & 0.1639 \\
26 & UZ & AS & 0.2154 & 66 & CD & AF & 0.1974 & 106 & YE & AS & 0.1902 & 146 & KR & AS & 0.1631 \\
27 & RE & AF & 0.2153 & 67 & NL & EU & 0.1973 & 107 & PY & SA & 0.1901 & 147 & CO & SA & 0.1618 \\
28 & IS & EU & 0.2137 & 68 & SG & AS & 0.1971 & 108 & CU & NA & 0.1900 & 148 & VN & AS & 0.1599 \\
29 & BA & EU & 0.2123 & 69 & SO & AF & 0.1967 & 109 & CM & AF & 0.1899 & 149 & JP & AS & 0.1499 \\
30 & MT & EU & 0.2116 & 70 & LB & AS & 0.1966 & 110 & LK & AS & 0.1897 & 150 & TW & AS & 0.1308 \\
31 & LA & AS & 0.2113 & 71 & TG & AF & 0.1963 & 111 & OM & AS & 0.1895 &  &  &  &  \\
32 & MQ & NA & 0.2107 & 72 & AE & AS & 0.1962 & 112 & FR & EU & 0.1891 &  &  &  &  \\
33 & NZ & OC & 0.2106 & 73 & IL & AS & 0.1958 & 113 & MY & AS & 0.1889 &  &  &  &  \\
34 & CH & EU & 0.2101 & 74 & SD & AF & 0.1956 & 114 & DO & NA & 0.1887 &  &  &  &  \\
35 & SE & EU & 0.2097 & 75 & NP & AS & 0.1956 & 115 & SA & AS & 0.1887 &  &  &  &  \\
36 & GP & NA & 0.2096 & 76 & ZA & AF & 0.1956 & 116 & PL & EU & 0.1884 &  &  &  &  \\
37 & US & NA & 0.2096 & 77 & CA & NA & 0.1953 & 117 & MA & AF & 0.1879 &  &  &  &  \\
38 & MU & AF & 0.2084 & 78 & ZW & AF & 0.1953 & 118 & MZ & AF & 0.1874 &  &  &  &  \\
39 & MM & AS & 0.2077 & 79 & KH & AS & 0.1952 & 119 & RW & AF & 0.1870 &  &  &  &  \\
40 & NO & EU & 0.2074 & 80 & PG & OC & 0.1949 & 120 & KE & AF & 0.1868 &  &  &  &  \\
\bottomrule
\end{tabular}
\caption{Country x CA Centralization Scores}
\label{tab:country_ca_all_scores}
\end{table}

\begin{table}[h!]
\begin{tabular}{llll|llll|llll|llll}
    \toprule
\multicolumn{1}{c}{\textbf{Rank}} & \multicolumn{2}{c}{\textbf{Country}} & \multicolumn{1}{c}{\textbf{CS}} & \multicolumn{1}{c}{\textbf{Rank}} & \multicolumn{2}{c}{\textbf{Country}} & \multicolumn{1}{c}{\textbf{CS}} & \multicolumn{1}{c}{\textbf{Rank}} & \multicolumn{2}{c}{\textbf{Country}} & \multicolumn{1}{c}{\textbf{CS}} & \multicolumn{1}{c}{\textbf{Rank}} & \multicolumn{2}{c}{\textbf{Country}} & \multicolumn{1}{c}{\textbf{CS}} \\
\midrule
1 & US & NA & 0.5853 & 41 & LY & AF & 0.3610 & 81 & TG & AF & 0.3284 & 121 & HR & EU & 0.2878 \\
2 & PR & NA & 0.5358 & 42 & MV & AS & 0.3609 & 82 & NL & EU & 0.3270 & 122 & AL & EU & 0.2781 \\
3 & TT & NA & 0.4821 & 43 & GH & AF & 0.3609 & 83 & SE & EU & 0.3258 & 123 & PY & SA & 0.2700 \\
4 & JM & NA & 0.4771 & 44 & SD & AF & 0.3608 & 84 & MG & AF & 0.3254 & 124 & EE & EU & 0.2694 \\
5 & CZ & EU & 0.4656 & 45 & BW & AF & 0.3600 & 85 & DZ & AF & 0.3252 & 125 & MN & AS & 0.2624 \\
6 & HU & EU & 0.4450 & 46 & ML & AF & 0.3595 & 86 & IN & AS & 0.3250 & 126 & AO & AF & 0.2592 \\
7 & PL & EU & 0.4265 & 47 & GT & NA & 0.3595 & 87 & AE & AS & 0.3245 & 127 & BE & EU & 0.2573 \\
8 & TH & AS & 0.4108 & 48 & NA & AF & 0.3591 & 88 & ZW & AF & 0.3233 & 128 & MK & EU & 0.2560 \\
9 & GR & EU & 0.4044 & 49 & ET & AF & 0.3586 & 89 & MO & AS & 0.3227 & 129 & MZ & AF & 0.2524 \\
10 & CR & NA & 0.4022 & 50 & IQ & AS & 0.3579 & 90 & HK & AS & 0.3223 & 130 & VN & AS & 0.2506 \\
11 & CA & NA & 0.4008 & 51 & GP & NA & 0.3552 & 91 & BD & AS & 0.3214 & 131 & CY & AS & 0.2486 \\
12 & BN & AS & 0.3979 & 52 & MQ & NA & 0.3539 & 92 & MU & AF & 0.3203 & 132 & UA & EU & 0.2470 \\
13 & PA & NA & 0.3951 & 53 & SY & AS & 0.3535 & 93 & BJ & AF & 0.3200 & 133 & LV & EU & 0.2421 \\
14 & MM & AS & 0.3945 & 54 & MT & EU & 0.3530 & 94 & LT & EU & 0.3186 & 134 & IS & EU & 0.2367 \\
15 & LA & AS & 0.3903 & 55 & AU & OC & 0.3530 & 95 & SG & AS & 0.3174 & 135 & CH & EU & 0.2356 \\
16 & BR & SA & 0.3856 & 56 & BF & AF & 0.3521 & 96 & SN & AF & 0.3166 & 136 & BY & EU & 0.2289 \\
17 & EG & AF & 0.3846 & 57 & DO & NA & 0.3517 & 97 & EC & SA & 0.3144 & 137 & ID & AS & 0.2272 \\
18 & HN & NA & 0.3837 & 58 & PH & AS & 0.3510 & 98 & ZA & AF & 0.3143 & 138 & BA & EU & 0.2228 \\
19 & RO & EU & 0.3811 & 59 & CL & SA & 0.3496 & 99 & AF & AS & 0.3142 & 139 & ME & EU & 0.2192 \\
20 & MW & AF & 0.3797 & 60 & FR & EU & 0.3481 & 100 & NP & AS & 0.3138 & 140 & TM & AS & 0.2128 \\
21 & TR & AS & 0.3776 & 61 & GB & EU & 0.3470 & 101 & CI & AF & 0.3128 & 141 & AT & EU & 0.2123 \\
22 & SK & EU & 0.3731 & 62 & VE & SA & 0.3469 & 102 & CD & AF & 0.3108 & 142 & AZ & AS & 0.2035 \\
23 & SO & AF & 0.3729 & 63 & GA & AF & 0.3468 & 103 & RE & AF & 0.3106 & 143 & GE & AS & 0.1936 \\
24 & NI & NA & 0.3723 & 64 & OM & AS & 0.3450 & 104 & NO & EU & 0.3098 & 144 & LU & EU & 0.1838 \\
25 & NG & AF & 0.3713 & 65 & RW & AF & 0.3439 & 105 & PE & SA & 0.3077 & 145 & AM & AS & 0.1794 \\
26 & SV & NA & 0.3701 & 66 & IR & AS & 0.3418 & 106 & BO & SA & 0.3076 & 146 & KZ & AS & 0.1629 \\
27 & JO & AS & 0.3701 & 67 & RU & EU & 0.3416 & 107 & MA & AF & 0.3055 & 147 & UZ & AS & 0.1569 \\
28 & IT & EU & 0.3700 & 68 & HT & NA & 0.3407 & 108 & TW & AS & 0.3054 & 148 & TJ & AS & 0.1526 \\
29 & KW & AS & 0.3699 & 69 & AR & SA & 0.3391 & 109 & BG & EU & 0.3051 & 149 & MD & EU & 0.1475 \\
30 & JP & AS & 0.3693 & 70 & NZ & OC & 0.3369 & 110 & SI & EU & 0.3043 & 150 & KG & AS & 0.1468 \\
31 & DK & EU & 0.3692 & 71 & CU & NA & 0.3367 & 111 & IE & EU & 0.3040 &  &  &  &  \\
32 & BH & AS & 0.3668 & 72 & CO & SA & 0.3364 & 112 & LK & AS & 0.3024 &  &  &  &  \\
33 & PG & OC & 0.3666 & 73 & ES & EU & 0.3355 & 113 & PK & AS & 0.3015 &  &  &  &  \\
34 & ZM & AF & 0.3658 & 74 & QA & AS & 0.3339 & 114 & PT & EU & 0.3009 &  &  &  &  \\
35 & LB & AS & 0.3647 & 75 & MX & NA & 0.3326 & 115 & IL & AS & 0.2971 &  &  &  &  \\
36 & FI & EU & 0.3646 & 76 & SA & AS & 0.3325 & 116 & UY & SA & 0.2966 &  &  &  &  \\
37 & UG & AF & 0.3635 & 77 & PS & AS & 0.3311 & 117 & DE & EU & 0.2920 &  &  &  &  \\
38 & YE & AS & 0.3620 & 78 & CM & AF & 0.3302 & 118 & RS & EU & 0.2914 &  &  &  &  \\
39 & KR & AS & 0.3613 & 79 & KE & AF & 0.3293 & 119 & MY & AS & 0.2905 &  &  &  &  \\
40 & KH & AS & 0.3610 & 80 & TZ & AF & 0.3284 & 120 & TN & AF & 0.2893 &  &  &  &  \\
\bottomrule
\end{tabular}
\caption{Country x TLD Centralization Scores}
\label{tab:country_tld_all_scores}
\end{table}
 

%% file: cs_all_countries.tex
\section{DNS, CA, and TLD Layers}
\label{appendix:cs_rest}

\subsection{Centralization continental trends}
\label{subappendix:continent_rest}

Figures~\ref{fig:centralization_dns_barplot},~\ref{fig:centralization_ca_barplot}, and~\ref{fig:centralization_tld_barplot} show the distribution of countries' centralization scores color-coded by continent for the DNS, CA, and TLD layers, respectively. See Figure~\ref{fig:centralization_barplot} in Section~\ref{subsec:hosting_overview} for corresponding results for hosting.

\csdnsbarplot

\cscabarplot

\cstldbarplot

\subsection{Cluster usage distribution}
\label{subappendix:stacked_rest}

Figures~\ref{fig:hosting_dns_stacked},~\ref{fig:hosting_ca_stacked}, and~\ref{fig:hosting_tld_stacked} show the category breakdowns of DNS, CA, and TLD use by country, sorted by countries' centralization scores. See Figure~\ref{fig:hosting_provider_stacked} in Section~\ref{subsec:hosting_categories} for corresponding results for hosting.

\dnsstackedbarplot
\vspace{20pt}

\castackedbarplot
\vspace{20pt}

\tldstackedbarplot

%% file: insularity_rest.tex
\section{Insularity continental trends}
\label{appendix:continent_rest_insularity}
Figures \ref{fig:insularity_barplot} and \ref{fig:insularity_dns_barplot} shows the distribution of countries' insularities color-coded by continent for hosting and DNS layers, respectively. See Figures \ref{fig:insularity_tld_barplot} and \ref{fig:insularity_ca_barplot} for corresponding results for TLDs and CAs

\insularityhostingbarplot
\insularitydnsbarplot